\documentclass[camera,letterpaper,nomarginnotes,nonarrowgutter]{jpaper}
\usepackage{fancyhdr}
\usepackage{balance}
\usepackage{cite}
\usepackage{amsmath,amssymb,amsfonts}
\usepackage{algorithmic}
\usepackage{graphicx}
\usepackage{textcomp}
\usepackage{xcolor}

\usepackage{tikzducks}
\usepackage{adjustbox}
\usepackage{multirow}
\usepackage{siunitx}
\usepackage{glossaries}
\usepackage[linesnumbered]{algorithm2e}
\usepackage{setspace}
\setkeys{glslink}{hyper=false}

\newcommand{\exploitingRowHammerAllCitations}[0]{\cite{fournaris2017exploiting,
poddebniak2018attacking, tatar2018throwhammer, carre2018openssl,
barenghi2018software, zhang2018triggering, bhattacharya2018advanced,
google-project-zero, kim2014flipping, rowhammergithub, seaborn2015exploiting,
van2016drammer, gruss2016rowhammer, razavi2016flip, pessl2016drama, xiao2016one,
bosman2016dedup, bhattacharya2016curious, burleson2016invited, qiao2016new,
brasser2017can, jang2017sgx, aga2017good, mutlu2017rowhammer,
tatar2018defeating, gruss2018another, lipp2018nethammer, van2018guardion,
frigo2018grand, cojocar2019eccploit,  ji2019pinpoint, mutlu2019rowhammer,
hong2019terminal, kwong2020rambleed, frigo2020trrespass, cojocar2020rowhammer,
weissman2020jackhammer, zhang2020pthammer, yao2020deephammer, deridder2021smash,
hassan2021utrr, jattke2022blacksmith, tol2022toward, kogler2022half,
orosa2024spyhammer, zhang2022implicit, liu2022generating, cohen2022hammerscope,
zheng2022trojvit, fahr2022frodo, tobah2022spechammer, rakin2022deepsteal,
aydin2022cyber, mus2022jolt, wang2022research,
lefforge2023reverse,fahr2022effects, kaur2022work, cai2022feasibility,
li2022cyberradar, roohi2022efficient, staudigl2022neurohammer, yang2022socially,
islam2022signature, mutlu2019retrospective,mutlu2023fundamentally,
luo2023rowpress, olgun2024read, lin2025gpuhammer, yuksel2025pudhammer, 
woo2025mitigations, bostanci2025understanding, meyer2026phoenix}}

\newcommand{\mitigatingRowHammerAllCitations}[0]{\cite{AppleRefInc,
rh-hp,rh-lenovo,greenfield2012throttling, kim2014flipping, kim2014architectural,
bains14d, bains14c, aweke2016anvil, bains-merged, son2017making,
seyedzadeh2018cbt,irazoqui2016mascat, you2019mrloc, lee2019twice,
park2020graphene, yaglikci2021security, yaglikci2021blockhammer,
frigo2020trrespass, kang2020cattwo, hassan2021utrr, qureshi2022hydra,
saileshwar2022randomized, brasser2017can, konoth2018zebram, van2018guardion,
vig2018rapid,  kim2022mithril, lee2021cryoguard, marazzi2022protrr,
zhang2022softtrr, joardar2022learning, juffinger2023csi, yaglikci2022hira,
saxena2022aqua, enomoto2022efficient, manzhosov2022revisiting, ajorpaz2022evax,
naseredini2022alarm, joardar2022machine, hassan2022case,
zhang2020leveraging,loughlin2021stop, devaux2021method, han2021surround,
fakhrzadehgan2022safeguard, saroiu2022price, saroiu2022configure,
loughlin2022moesiprime, zhou2022lt, hong2023dsac, mutlu2023fundamentally,
marazzi2023rega, di2023copy, sharma2022review, woo2023scalable, park2022row,
wi2023shadow, kim2023ddr5, gude2023defending, guha2022criticality,
france2022modeling, france2022reducing, bennett2021panopticon,
arikan2022processor, tomita2022extracting, saxena2023pt, zhou2023dnndefender,
woo2023rampart, kim2023how, olgun2024abacus, yaglikci2024spatial,
bostanci2024comet, saroiu2024ddr5, saxena2024start, jedecddr5c,
canpolat2024understanding,jaleel2024pride,saxena2024rubix, qureshi2026salt, 
taneja2026mirza, vittal2025mopac, kim2025per, fiedler2026memory}}

\newcommand{\refreshBasedRowHammerDefenseCitations}[0]{\cite{lee2019twice,
seyedzadeh2017cbt, seyedzadeh2018cbt, kang2020cattwo, park2020graphene,
kim2022mithril, kim2014architectural, bains2016row,
aweke2016anvil, jedecddr5c,olgun2024abacus,saxena2024start, bains14d,
yaglikci2024spatial,bostanci2024comet,qureshi2022hydra,yaglikci2022hira,
kim2014flipping,yaglikci2021security,
canpolat2024understanding,canpolat2025chronus,hassan2024self, 
canpolat2024breakhammer, qureshi2026salt, taneja2026mirza, vittal2025mopac, kim2025per}}

\newcommand{\readDisturbanceCharacterizationCitations}[0]{\cite{kim2014flipping,kim2020revisiting,orosa2021deeper,
yaglikci2022understanding,olgun2023hbm,lang2023blaster,
he2023whistleblower,yaglikci2024spatial,luo2023rowpress,olgun2024read,nam2024dramscope,nam2023xray, 
tugrul2025understanding, luo2025revisiting, olgun2025variable,hassan2021utrr,frigo2020trrespass,park2016experiments,yuksel2025columndisturb}}

\newcommand{\figref}[1]{Fig.~\ref{#1}}
\newcommand{\secref}[1]{§\ref{#1}}

\newcommand{\numDDRchips}[0]{\atbcr{1}{212}}

\newcommand{\nummodules}[0]{\atbcr{1}{28}}
\newcommand{\numtestedrows}[0]{\atbcr{1}{317985}}
\newcommand{\numRDTmeasurements}[0]{\atbcr{1}{1000}}

\newlength{\HeightReference}
\newlength{\DepthReference}
\newlength{\Width}%

\newcommand{\MyColorBox}[2][red]%
{%
    \settowidth{\Width}{#2}%
    \colorbox{#1}%
    {%
        \raisebox{-\DepthReference}%
        {%
                \parbox[b][\HeightReference+\DepthReference][c]{\Width}{\centering#2}%
        }%
    }%
}
\newcounter{obs}
\newcounter{take}
\newcommand\take[1]{%
   \refstepcounter{take}
   \vspace{1mm}
  \noindent
  \begin{tabular}{|p{0.95\linewidth}|}
       \hline
       \textbf{{Takeaway \thetake}.} {{#1}}\\
       \hline 
  \end{tabular}
}

\definecolor{iy}{rgb}{0.0, 0.5, 0.2}
\definecolor{nbc}{rgb}{0.5, 0.0, 0.13}
\definecolor{moegi}{rgb}{0.357, 0.537, 0.188}
\definecolor{amethyst}{rgb}{0.6, 0.4, 0.8}

\newif\ifdraft
\newif\ifsubmission
\newif\ifhpcarevision
\newif\ifshepherdcomments
\newif\ifcamerareadyiterations
\draftfalse
\submissiontrue
\hpcarevisionfalse
\shepherdcommentsfalse
\camerareadyiterationsfalse

\ifdraft
    \paperwidth=\dimexpr \paperwidth + 5cm\relax
    \oddsidemargin=\dimexpr\oddsidemargin + 2cm\relax
    \evensidemargin=\dimexpr\evensidemargin + 2cm\relax
    \marginparwidth=\dimexpr \marginparwidth + 2cm\relax
    \usepackage[colorinlistoftodos,prependcaption,textsize=small]{todonotes}

    \newcommand{\atbcomment}[1]{\todo[size=\scriptsize, linecolor=blue, bordercolor=blue, backgroundcolor=white]{\textcolor{blue}{\textbf{@atb:} #1}}}

    \newcommand{\ieycomment}[1]{\todo[size=\scriptsize, linecolor=iy, bordercolor=iy, backgroundcolor=white]{\textcolor{iy}{\textbf{@iey:} #1}}}
    \newcommand{\nbcomment}[1]{\todo[size=\scriptsize, linecolor=nbc, bordercolor=nbc, backgroundcolor=white]{\textcolor{nbc}{\textbf{@nb:} #1}}}
    \newcommand{\agycomment}[1]{\todo[size=\scriptsize, linecolor=nbc, bordercolor=nbc, backgroundcolor=white]{\textcolor{orange}{\textbf{@agy:} #1}}}
    \newcommand{\agyurgentcomment}[1]{\todo[size=\scriptsize, linecolor=nbc, bordercolor=nbc, backgroundcolor=white]{\textcolor{red}{\textbf{[!ALARM!]@agy:} #1}}}
    \newcommand{\gfcomment}[1]{\todo[size=\scriptsize, linecolor=blue, bordercolor=blue, backgroundcolor=white]{\textcolor{blue}{\textbf{@gf:} #1}}}
      
    \newcommand{\param}[1]{\textcolor{red}{#1}}
    \newcommand{\copied}[1]{\textcolor{gray}{#1}}
    \newcommand{\outline}[1]{\noindent\textcolor{orange}{\textbf{Outline:} \emph{#1}}}
    
    \newcommand{\nb}[1]{\textcolor{nbc}{#1}}
    \newcommand{\iey}[1]{\textcolor{iy}{#1}}
    
    \newcommand{\gf}[1]{\textcolor{blue}{#1}}
    \newcommand{\agy}[1]{\textcolor{orange}{#1}}

    \newcommand{\gra}[1]{\textcolor{amethyst}{#1}}

    \newcommand{\hpcarevcommon}[1]{#1}
    \newcommand{\hpcareva}[1]{#1}
    \newcommand{\hpcarevb}[1]{#1}
    \newcommand{\hpcarevc}[1]{#1}
    
    \newcommand{\hpcareve}[1]{#1}

    \newcommand{\hpcalabel}[1]{}

    \newcommand{\atbcr}[2]{#2}
    \newcommand{\omcr}[2]{#2}
    \newcommand{\atbcrcomment}[2]{}
    \newcommand{\omcrcomment}[2]{}
\else
\ifsubmission
    \newcommand{\atbcomment}[1]{}
    \newcommand{\ieycomment}[1]{}
    \newcommand{\nbcomment}[1]{}
    \newcommand{\agycomment}[1]{}
    \newcommand{\agyurgentcomment}[1]{}
    \newcommand{\gfcomment}[1]{}

    \newcommand{\param}[1]{#1}
    \newcommand{\copied}[1]{#1}
    \newcommand{\outline}[1]{}
    
    \newcommand{\iey}[1]{#1}
    \newcommand{\nb}[1]{#1}
    
    \newcommand{\gf}[1]{#1}
    \newcommand{\agy}[1]{{#1}}

    \newcommand{\gra}[1]{#1}

    \newcommand{\hpcarevcommon}[1]{#1}
    \newcommand{\hpcareva}[1]{#1}
    \newcommand{\hpcarevb}[1]{#1}
    \newcommand{\hpcarevc}[1]{#1}
    
    \newcommand{\hpcareve}[1]{#1}

    \newcommand{\hpcalabel}[1]{}

    \newcommand{\atbcr}[2]{#2}
    \newcommand{\omcr}[2]{#2}
    \newcommand{\atbcrcomment}[2]{}
    \newcommand{\omcrcomment}[2]{}
\else
\ifhpcarevision
    \usepackage[colorinlistoftodos,prependcaption,textsize=small]{todonotes}

    \newcommand{\atbcomment}[1]{}
    \newcommand{\ieycomment}[1]{}
    \newcommand{\nbcomment}[1]{}
    \newcommand{\agycomment}[1]{}
    \newcommand{\agyurgentcomment}[1]{}
    \newcommand{\gfcomment}[1]{}

    \newcommand{\param}[1]{#1}
    \newcommand{\copied}[1]{#1}
    \newcommand{\outline}[1]{}
    
    \newcommand{\iey}[1]{#1}
    \newcommand{\nb}[1]{#1}
    
    \newcommand{\gf}[1]{#1}
    \newcommand{\agy}[1]{{#1}}

    \newcommand{\gra}[1]{#1}

    \newcommand{\hpcarevcommon}[1]{\textcolor{blue}{#1}}
    \newcommand{\hpcareva}[1]{\textcolor{red}{#1}}
    \newcommand{\hpcarevb}[1]{\textcolor{moegi}{#1}}
    \newcommand{\hpcarevc}[1]{\textcolor{orange}{#1}}
    
    \newcommand{\hpcareve}[1]{\textcolor{cyan}{#1}}

    \definecolor{babyblueeyes}{rgb}{0.63, 0.79, 0.95} 
    \newcommand{\hpcalabel}[1]{\todo[size=\scriptsize, linecolor=black, bordercolor=black, backgroundcolor=babyblueeyes]{#1}}

    \newcommand{\atbcr}[2]{#2}
    \newcommand{\omcr}[2]{#2}
    \newcommand{\atbcrcomment}[2]{}
    \newcommand{\omcrcomment}[2]{}
\else
\ifshepherdcomments
    \paperwidth=\dimexpr \paperwidth + 5cm\relax
    \oddsidemargin=\dimexpr\oddsidemargin + 2cm\relax
    \evensidemargin=\dimexpr\evensidemargin + 2cm\relax
    \marginparwidth=\dimexpr \marginparwidth + 2cm\relax
    \usepackage[colorinlistoftodos,prependcaption,textsize=small]{todonotes}

    \newcommand{\atbcomment}[1]{}
    \newcommand{\ieycomment}[1]{}
    \newcommand{\nbcomment}[1]{}
    \newcommand{\agycomment}[1]{}
    \newcommand{\agyurgentcomment}[1]{}
    \newcommand{\gfcomment}[1]{}

    \newcommand{\param}[1]{#1}
    \newcommand{\copied}[1]{#1}
    \newcommand{\outline}[1]{}
    
    \newcommand{\iey}[1]{#1}
    \newcommand{\nb}[1]{#1}
    
    \newcommand{\gf}[1]{#1}
    \newcommand{\agy}[1]{{#1}}

    \newcommand{\gra}[1]{#1}

    \newcommand{\hpcarevcommon}[1]{#1}
    \newcommand{\hpcareva}[1]{#1}
    \newcommand{\hpcarevb}[1]{#1}
    \newcommand{\hpcarevc}[1]{#1}
    
    \newcommand{\hpcareve}[1]{#1}

    \definecolor{babyblueeyes}{rgb}{0.63, 0.79, 0.95} 
    \newcommand{\hpcalabel}[1]{}

    \newcommand{\atbcr}[2]{#2}
    \newcommand{\omcr}[2]{#2}
    \newcommand{\atbcrcomment}[2]{}
    \newcommand{\omcrcomment}[2]{}
\else
\ifcamerareadyiterations
    \paperwidth=\dimexpr \paperwidth + 5cm\relax
    \oddsidemargin=\dimexpr\oddsidemargin + 2cm\relax
    \evensidemargin=\dimexpr\evensidemargin + 2cm\relax
    \marginparwidth=\dimexpr \marginparwidth + 2cm\relax
    \usepackage[colorinlistoftodos,prependcaption,textsize=small]{todonotes}

    \newcommand{\atbcomment}[1]{}
    \newcommand{\ieycomment}[1]{}
    \newcommand{\nbcomment}[1]{}
    \newcommand{\agycomment}[1]{}
    \newcommand{\agyurgentcomment}[1]{}
    \newcommand{\gfcomment}[1]{}

    \newcommand{\param}[1]{#1}
    \newcommand{\copied}[1]{#1}
    \newcommand{\outline}[1]{}
    
    \newcommand{\iey}[1]{#1}
    \newcommand{\nb}[1]{#1}
    
    \newcommand{\gf}[1]{#1}
    \newcommand{\agy}[1]{{#1}}

    \newcommand{\gra}[1]{#1}

    \newcommand{\hpcarevcommon}[1]{#1}
    \newcommand{\hpcareva}[1]{#1}
    \newcommand{\hpcarevb}[1]{#1}
    \newcommand{\hpcarevc}[1]{#1}
    
    \newcommand{\hpcareve}[1]{#1}

    \definecolor{babyblueeyes}{rgb}{0.63, 0.79, 0.95} 
    \newcommand{\hpcalabel}[1]{}

    \newcommand{\atbcr}[2]{\ifnum#1=\value{version}\textcolor{red}{#2}\else{#2}\fi}
    \newcommand{\omcr}[2]{\ifnum#1=\value{version}\textcolor{blue}{#2}\else{#2}\fi}
    \newcommand{\atbcrcomment}[2]{\ifnum#1=\value{version}\todo[size=\scriptsize, linecolor=orange, bordercolor=orange, backgroundcolor=white]{\textcolor{red}{Atb:~#2}}\else{}\fi}
    \newcommand{\omcrcomment}[2]{\ifnum#1=\value{version}\todo[size=\scriptsize, linecolor=orange, bordercolor=orange, backgroundcolor=white]{\textcolor{blue}{Onur:~#2}}\else{}\fi}
\fi
\fi
\fi
\fi
\fi

\def\UrlBreaks{\do\/\do-\/\do.\/\do:}

\expandafter\def\expandafter\UrlBreaks\expandafter{\UrlBreaks
  \do\a\do\b\do\c\do\d\do\e\do\f\do\g\do\h\do\i\do\j
  \do\k\do\l\do\m\do\n\do\o\do\p\do\q\do\r\do\s\do\t
  \do\u\do\v\do\w\do\x\do\y\do\z\do\A\do\B\do\C\do\D
  \do\E\do\F\do\G\do\H\do\I\do\J\do\K\do\L\do\M\do\N
  \do\O\do\P\do\Q\do\R\do\S\do\T\do\U\do\V\do\W\do\X
  \do\Y\do\Z}

\newcommand{\hcfirst}[0]{HC_{first}}
\newacronym{hcfirst}{$\hcfirst$}{the minimum {hammer count} {required to induce the first bitflip}} 
\newacronym{ber}{$BER$}{bit error rate}
\newacronym{wcdp}{$WCDP$}{worst-case data pattern}

\newacronym{taggon}{$t_{AggOn}$}{aggressor row on time}
\newacronym{taggoff}{$t_{AggOff}$}{time that an aggressor row stays closed}
\newacronym{tras}{$t_{RAS}$}{charge restoration latency}
\newacronym{trp}{$t_{RP}$}{precharge latency}
\newcommand{\trc}[0]{t_{RC}}
\newacronym{trc}{$\trc{}$}{row activation cycle}
\newacronym{trcd}{$t_{RCD}$}{row activation latency}
\newacronym{tcl}{$t_{CL}$}{column access latency}
\newacronym{tcwl}{$t_{CWL}$}{column write latency}
\newcommand{\tfaw}[0]{t_{FAW}}
\newacronym{tfaw}{$\tfaw{}$}{four row activation window}
\newcommand{\trefw}[0]{t_{REFW}}
\newacronym{trefw}{$\trefw{}$}{refresh window}
\newcommand{\trefi}[0]{t_{REFI}}
\newacronym{trefi}{$\trefi{}$}{refresh interval}
\newacronym{trrslack}{$t_{RefSlack}$}{{the maximum delay between the time a {periodic}/{preventive} refresh is generated and the time the refresh is performed}}
\newacronym{tapa}{$t_{APA}$}{the latency of issuing $ACT-PRE-ACT$ command sequence}
\newacronym{ref}{$REF$}{refresh}
\newacronym{act}{$ACT$}{activate}
\newacronym{pre}{$PRE$}{precharge}
\newcommand{\trfc}[0]{t_{RFC}}
\newacronym{trfc}{$\trfc{}$}{refresh latency}
\newacronym{iqr}{$IQR$}{interquartile range}
\newacronym{cv}{$CV$}{the coefficient of variation}
\newacronym{hc}{$HC$}{hammer count}
\newcommand{\pth}[0]{p_{th}}
\newacronym{pth}{$\pth{}$}{{PARA's probability threshold}}
\newcommand{\pf}[0]{p_{failure}}
\newacronym{pf}{$\pf{}$}{failure probability over a sufficiently long time}
\newcommand{\prh}[0]{p_{RH}}
\newacronym{prh}{$\prh{}$}{reliability target for a \gls{trefw}}

\newcommand{\cchip}[0]{D_{chip}}
\newacronym{cchip}{$\cchip{}$}{chip density}

\newcommand{\rbcpki}[0]{RBCPKI}
\newacronym{rbcpki}{$\rbcpki{}$}{row buffer conflicts per kilo instruction}

\newcommand{\mpki}[0]{MPKI}
\newacronym{mpki}{$\mpki{}$}{misses per kilo instruction}

\newcommand{\vdd}[0]{V_{DD}}
\newacronym{vdd}{$\vdd{}$}{supply voltage}

\newcommand{\gnd}[0]{GND}
\newacronym{gnd}{$\gnd{}$}{ground}

\newcommand{\rd}[0]{RD}
\newacronym{rd}{$\rd{}$}{read}

\newcommand{\dramwr}[0]{WR}
\newacronym{wr}{$\dramwr{}$}{write}

\newacronym{minrdt}{$RDT_{min}$}{smallest RDT value across all tested rows}

\newcommand{\X}[0]{DiscoRD}

\def\BibTeX{{\rm B\kern-.05em{\sc i\kern-.025em b}\kern-.08em
    T\kern-.1667em\lower.7ex\hbox{E}\kern-.125emX}}
\begin{document}

\pdfpagewidth=8.5in
\pdfpageheight=11in

\newcommand{\iscasubmissionnumber}{2076}

\pagenumbering{arabic}

\title{\LARGE{DiscoRD: An Experimental Methodology for Quickly Discovering\\the Reliable Read Disturbance Threshold of Real DRAM Chips}}

\author{Ataberk Olgun$\dagger$ \quad
F. Nisa Bostanc{\i}$\dagger$ \quad 
\.{I}smail Emir Y\"{u}ksel$\dagger$ \quad 
Haocong Luo$\dagger$ \quad 
\\
Minesh Patel$\ddagger$ \quad 
A. Giray Ya\u{g}l{\i}k\c{c}{\i}$\dagger$ \quad 
Onur Mutlu$\dagger$\vspace{-3mm}\\\\
\emph{ETH Zurich}$\dagger$ \qquad \emph{Rutgers University}$\ddagger$}

\maketitle
\makeatletter
\g@addto@macro{\normalsize}{%
  \setlength{\abovedisplayskip}{1pt plus 0.5pt minus 1pt}
  \setlength{\belowdisplayskip}{1pt plus 0.5pt minus 1pt}
  \setlength{\abovedisplayshortskip}{0pt}
  \setlength{\belowdisplayshortskip}{0pt}
  \setlength{\intextsep}{1pt plus 1pt minus 1pt}
  \setlength{\textfloatsep}{1pt plus 1pt minus 1pt}
  \setlength{\skip\footins}{1pt plus 1pt minus 1pt}
  \setlength{\abovecaptionskip}{1pt plus 0pt minus 0pt}}
\makeatother

\makeatletter
\g@addto@macro{\normalsize}{%
 \setlength{\abovedisplayskip}{0pt plus 1pt minus 1pt}
 \setlength{\belowdisplayskip}{0pt plus 1pt minus 1pt}
  \setlength{\abovedisplayshortskip}{0pt}
  \setlength{\belowdisplayshortskip}{0pt}
  \setlength{\intextsep}{0pt plus 1pt minus 1pt}
  \setlength{\textfloatsep}{1pt plus 1pt minus 1pt}
  \setlength{\skip\footins}{2pt plus 1pt minus 1pt}}
   \setlength{\abovecaptionskip}{0pt plus 1pt minus 1pt}
   \setlength{\belowcaptionskip}{0pt plus 1pt minus 1pt}
\makeatother

\thispagestyle{plain}
\pagestyle{plain}


\begin{abstract}
Modern DRAM chips suffer from read disturbance. Repeatedly activating an
\emph{aggressor DRAM row} (RowHammer) or keeping an aggressor row open for a
very long time (RowPress) induces read disturbance bitflips in a physically
nearby \emph{victim DRAM row}\ieycomment{Why single row?}. State-of-the-art read disturbance mitigations
rely on the \emph{read disturbance threshold} (RDT) (e.g., the number of
aggressor row activations needed to induce the first read disturbance bitflip)
to securely and performance- and energy-efficiently prevent read disturbance
bitflips\ieycomment{I would put "to securely..." part to the beginning of this sentence.}. However, accurately and exhaustively\ieycomment{If you accurately and exhaustively a bit weird. To accurately identify the RDT you need to exhaustively characterize. not that important for now} characterizing the RDT of every
DRAM row in a chip is time intensive. 

Rapidly determining RDT is important for enabling secure, performance- and
energy-efficient systems. Our goal is to develop and evaluate a reliable and
rapid read disturbance testing methodology. To that end, we develop \X{}
building on the key results of an extensive experimental characterization study
using \numDDRchips{} real DDR4 chips whereby we measure the RDT of hundreds of
thousands of DRAM rows millions of times. Two important empirical
observations drive \X{}. First, the \gls{minrdt} significantly changes over
time, by at least \param{21\%}. Second, measuring \gls{minrdt} once and using it
as the RDT for all rows can lead to multiple read disturbance bitflips across
the DRAM chip\ieycomment{maybe also good to quantify}. Thus, we posit that a practical and secure approach to mitigating
read disturbance bitflips requires error tolerance and a means for quantitatively evaluating the security guarantees provided by such error tolerance. 

We develop an empirical model for read disturbance bitflips and evaluate the
probability of read-disturbance-induced uncorrectable errors when a read
disturbance mechanism is configured using a \emph{single} \gls{minrdt}
measurement. Using this model we demonstrate that 1)~relying on a lightweight
error-correcting code (ECC) alone yields relatively high uncorrectable error
probability (e.g., as many as one such error approximately every \param{1.4e+04}
hours) and 2)~combining ECC, infrequent memory scrubbing, and configurable read
disturbance mitigation mechanisms can greatly reduce the error probability
(e.g., to one error approximately every \param{1.8e+07} hours). Building on our
observations and analyses, we discuss how to more precisely identify the RDT of
each individual row. Our results show that error tolerance, memory scrubbing,
online profiling, and run-time configurable read disturbance mitigation
techniques are important to enable secure and energy-efficient spatial-variation
aware read disturbance mitigations.\ieycomment{You also have svard results maybe
add that as well? For example,...} We hope that \X{} drives research that
enables us to quantitatively navigate the performance/cost -- reliability
tradeoff space for read disturbance mitigation techniques.

\end{abstract}

\section{Introduction}
\label{sec:introduction}

\copied{Read disturbance~\cite{kim2014flipping,
mutlu2019retrospective,mutlu2023fundamentally,luo2023rowpress, olgun2024read,
mutlu2017rowhammer} (e.g., RowHammer and RowPress) in modern DRAM chips is a
widespread phenomenon and is reliably used for breaking memory
isolation~\exploitingRowHammerAllCitations{}, a fundamental building block for
building robust systems.} \copied{Repeatedly opening/activating and closing a
DRAM row (i.e., aggressor row) \emph{many times} (e.g., tens of thousands of
times) induces \emph{RowHammer bitflips} in physically nearby rows (i.e., victim
rows)~\cite{kim2014flipping}. Keeping the aggressor row open for a long period
of time amplifies the effects of read disturbance and induces \emph{RowPress
bitflips}, \emph{without} {requiring} \emph{many} repeated aggressor row
activations~\cite{luo2023rowpress}. Read disturbance mitigation
techniques~\mitigatingRowHammerAllCitations{} prevent read disturbance bitflips
by \emph{preventively} refreshing (i.e., opening and closing) a victim row
\emph{before a bitflip manifests in that row}.}

Accurately identifying the \emph{amount of read disturbance} that a victim row
can withstand before experiencing a bitflip (called the read disturbance
threshold~\cite{olgun2025variable,olgun2024read}, or RDT) is important
to securely mitigate read disturbance errors. However, accurately identifying
the RDT of every DRAM row (billions of such rows) in a DRAM chip may require
tens of thousands of RDT measurements because RDT changes significantly and
unpredictably over time, as described by the Variable Read Disturbance (VRD)
phenomenon~\cite{olgun2025variable}. 
Identifying the RDT of
every DRAM row is time intensive because state-of-the-art integrated circuit
test times are measured in seconds to minutes~\cite{ieee2019heterogeneous,
vandegoor2004industrial}, while measuring the RDT thousands of times (for all DRAM rows
in one chip under various environmental conditions) can require months of testing~\cite{olgun2025variable}.

While identifying the RDT of every DRAM row in a chip is time and energy
intensive, identifying a reliable, minimum RDT value that is a lower bound for
the RDTs of all DRAM rows in a chip\footnote{The read disturbance threshold
values are \emph{not} specified by the standard DRAM specifications and DRAM
manufacturers do \emph{not} disclose the read disturbance threshold value for
their DRAM chips. Thus, system designers must determine the read disturbance
threshold by extensively testing DRAM chips for read disturbance bitflips.} is
1)~sufficient to securely, albeit performance- and energy-inefficiently mitigate
read disturbance errors and 2)~could be done with fewer RDT measurements than it
would take to identify the RDT of every DRAM row. A seemingly straightforward
(and \emph{not} very time intensive), \emph{one size fits all} approach to
identifying this lower bound RDT value is to 1)~measure the RDT of every DRAM
row once, 2)~find the minimum RDT across all rows, and 3)~use a large safety
margin for the minimum RDT to cover all read disturbance bitflips. 
There are two
important problems with this approach.
\hpcalabel{Common Concern \#3}First, \hpcarevcommon{the literature lacks} an
established read disturbance testing methodology. \hpcarevcommon{Thus,} the
performance of the one size fits all approach is \emph{not} evaluated by any
prior work. For example, we do \emph{not} know the tradeoff between the RDT
safety margin and the failure probability (e.g., probability of a read
disturbance bitflip). Making definitive conclusions about the effectiveness of
guardbands is extremely challenging because doing so requires large-scale
experimental characterization. Second, a large fraction of DRAM cells have
higher read disturbance thresholds than the most read-disturbance-vulnerable
cell in a chip~\cite{yaglikci2024spatial}. Therefore, read disturbance
mitigation mechanisms configured using the one size fits all approach would
cause significant energy waste and degrade system performance by unnecessarily
refreshing potential victim rows. \hpcalabel{Common Concern
\#4}\hpcarevcommon{Numerous experimental studies
(e.g.,~\cite{kim2020revisiting,luo2023rowpress,cojocar2020rowhammer,frigo2020trrespass})
demonstrate that RDT reduces in newer generation DRAM chips. Therefore,
determining a reliable RDT value (that is neither too high to cause errors nor
too low to unnecessarily induce performance overheads) and
spatial-variation-aware, adaptive read disturbance mitigations are both very
important for continued DRAM density scaling without jeopardizing system
dependability and performance.}

Our \textbf{goal} is to enable quantitative reasoning about the RDT profiling
time and the failure (e.g., manifestation of a read disturbance bitflip)
probability design space. To that end, we develop the first reliable and rapid
RDT testing methodology, \X{}, that quickly identifies the minimum RDT across
all DRAM rows in a chip with a single measurement. \hpcarevcommon{\X{}
comprises} an empirical RDT model\hpcarevcommon{, that we build by
experimentally characterizing DRAM chips,} to quantify the read disturbance
bitflip probability when the identified RDT value is used for a read disturbance
mitigation technique. \hpcalabel{Common Concern \#3-\#4}\hpcarevcommon{By doing
so, \X{} enables designers to quantitatively reason about the security
guarantees of their read disturbance mitigation based on rapidly-gathered
empirical results.} To construct and evaluate \X{}, we rigorously experimentally
characterize \numDDRchips{} real DDR4 
DRAM chips for read disturbance bitflips, measuring the RDT of
\param{\numtestedrows{}} rows \param{\numRDTmeasurements{}} times. 

\glsresetall{}

We perform two case studies using \X{}. First, we evaluate the reliability of
the \emph{one size fits all} RDT testing approach. Second, \hpcalabel{Common
Concern \#3}\hpcarevcommon{we perform the first combined analysis taking both
spatial and temporal distribution in RDT into account and} evaluate the
performance of an RDT testing approach \hpcarevcommon{for enabling} 
performance-efficient read disturbance mitigation techniques that
account for the spatial variation in RDT across rows in a DRAM chip to reduce
the overheads of unnecessary read disturbance mitigative actions (i.e.,
Svärd~\cite{yaglikci2024spatial}).

\noindent
\textbf{\emph{One size fits all} RDT Testing.} Based on our extensive empirical
characterization results, we find that 1) a
\param{21.0}\% safety margin applied to the \gls{minrdt} prevents all observed
read disturbance bitflips over the course of 1K successive RDT measurements, 2)
up to \param{12} DRAM rows in a DRAM chip can exhibit read disturbance bitflips
if \gls{minrdt} is used as a read disturbance threshold for all DRAM rows, and
3) the DRAM row that yields \gls{minrdt} in one measurement can have a very
large RDT value (e.g., become the row with the \param{213th} smallest RDT across
all tested rows in a bank) in another measurement. Therefore, identifying all
potential DRAM rows that can fall below an \gls{minrdt} value (e.g., with the
goal of disabling such rows to prevent all read disturbance bitflips), even for
very low read disturbance values is challenging. We posit from our experimental
analysis that a practical and secure approach to mitigating read disturbance
bitflips likely requires some error tolerance\hpcalabel{E4}\hpcareve{. One} read
disturbance threshold measurement is unlikely to yield the minimum \gls{minrdt}
across time.

To quantify the effectiveness of combining error tolerance with read disturbance
mitigation in the presence of temporal variation in RDT, we use an empirical
data-driven model for read disturbance bitflips. \hpcarevcommon{We make two key
observations. First,} we show that the probability of uncorrectable error (when
a lightweight single error-correcting code that can correct a single bitflip in
a 136-bit codeword~\cite{mineshphd, alam2022comet, kim2023unity} is used for
error tolerance) significantly varies between error models constructed to
represent each tested module. Our model predicts that an uncorrectable error can
happen as rapidly as every \param{1.42e+04} and as slowly as every
\param{1.02e+05} hours. \hpcarevcommon{Second,} the error probability can be
significantly curbed (and mean time between errors can be increased) using
infrequent memory scrubbing and configurable read disturbance thresholds for
every DRAM row: 1)~checking for and correcting errors in every row in a DRAM
bank with an approximately 65-hour period and 2)~applying a very large safety
margin to the RDT of a DRAM row after it is found to exhibit bitflips can
improve the worst-case predicted mean time between uncorrectable errors from
\param{1.42e+04} hours to \param{1.84e+07} hours.

\noindent
\textbf{\hpcarevcommon{Spatial and Temporal Distribution in RDT Analysis} for
Svärd~\cite{yaglikci2024spatial}.} Selectively applying a large safety margin to
the RDT of a DRAM row that exhibits read disturbance bitflips requires a
configurable read disturbance mitigation substrate. This substrate is already
provided by Svärd~\cite{yaglikci2024spatial} to exploit the spatial variation in
RDT across rows in a bank. We experimentally study the read disturbance bitflips
that manifest when a single RDT measurement is used for the large majority
(approximately 90\%, $RDT_{90\%}$) of the DRAM rows. We highlight two important
observations. First, using $RDT_{90\%}$ as the RDT for all DRAM rows yields many
read disturbance bitflips (up to \param{2437} in all tested rows in a module).
Second, a large fraction of these bitflips \emph{cannot} be identified with a
single RDT measurement (e.g., \param{48.1\%} of rows in one tested module).
\hpcalabel{E4}\hpcareve{We make the key takeaway
that a spatial variation-aware approach to read disturbance requires multiple
rounds of read disturbance testing for dependable system operation.} 

\hpcalabel{Common Concern \#4 and E4}\hpcarevcommon{Using \X{},} we evaluate the
uncorrectable error probability for Svärd assuming the system designer performs
few tens of test repetitions to uncover a large fraction of
temporal-variation-induced bitflips. \hpcarevcommon{We show that 1)~Svärd can be
configured to match the error probability of the \emph{one size fits all}
approach using \X{} (e.g., with an estimated mean time to uncorrectable error of
\param{7.25e+06} hours), and 2)~such a configuration of Svärd improves
performance compared to the \emph{one size fits all} approach (e.g., by
\param{32}\% for PARA~\cite{kim2014flipping} and \param{8}\% for
Chronus~\cite{canpolat2025chronus} on average across 60 diverse workloads).}

We make the following contributions:
\begin{itemize}
    \item We develop the first reliable read 
    disturbance threshold testing methodology, \X{}. \X{} enables
    quantitatively reasoning about the read disturbance testing time -- 
    read disturbance bitflip probability tradeoff.
    \item We rigorously characterize \numDDRchips{} real DDR4 DRAM chips 
    for read disturbance bitflips. Our results strongly suggest that
    a secure read disturbance mitigation technique should employ 
    some form of error tolerance.
    \item We provide an empirical model for read disturbance bitflips to evaluate 
    the probability of uncorrectable errors. We demonstrate that 
    using a lightweight error-correcting code (ECC) alone yields a 
    high uncorrectable error probability and that 
    combining ECC with other techniques can significantly reduce this probability.
    \item We adapt \X{} to account RDT's spatial variation. Using this adaptation
    we show that a spatial variation-aware read disturbance mitigation technique 
    must also be aware of temporal-variation in RDT to securely prevent bitflips. 
    \item We quantify 
    the uncorrectable error probability and system performance 
    improvements for a spatial and temporal variation-aware read disturbance mitigation technique. 
\end{itemize}

\section{Background and Motivation}
\copied{This section provides a concise overview of 1)~DRAM organization, 2)~DRAM
operation\gra{,} 3)~DRAM read disturbance, \omcr{2}{and 4)~our motivation in
this work}.}

\subsection{DRAM Organization}
\begin{figure}[!ht]
    \centering
    \includegraphics[width=\linewidth]{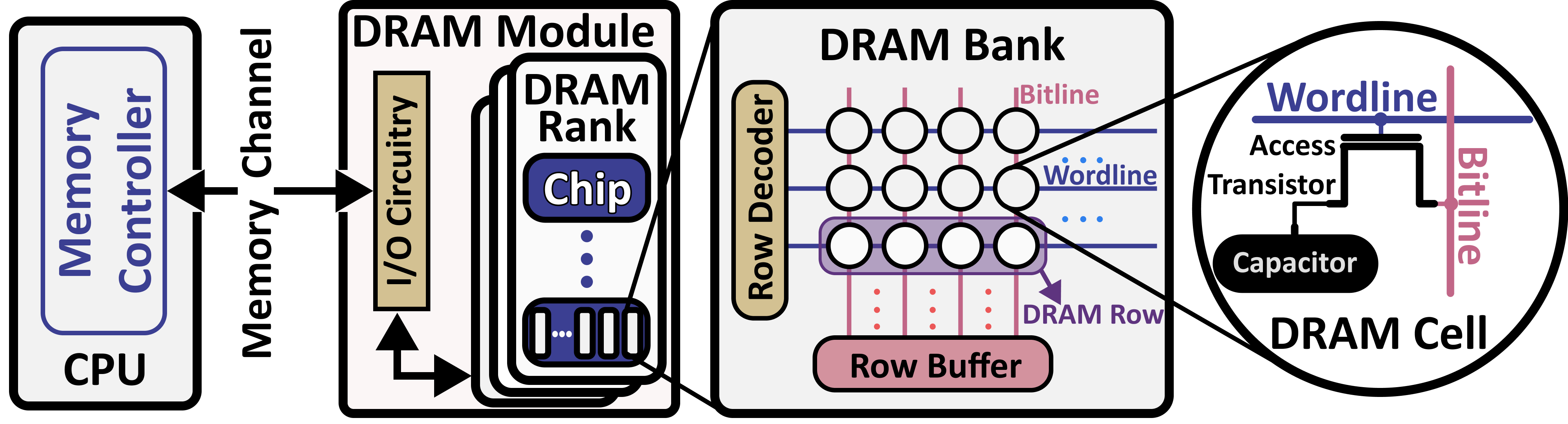}
    \caption{DRAM module, rank, chip, and bank organization}
    \label{fig:dram_organization}
\end{figure}

\copied{\figref{fig:dram_organization} shows the organization of a DRAM-based
memory system. A \iey{\emph{memory channel}} connects the processor (CPU) to
\iey{a \emph{DRAM module}, where a module consists of multiple \emph{DRAM
ranks}. A DRAM rank is formed by a set of \emph{DRAM chips} that are operated in
lockstep}. Each \iey{DRAM} chip has multiple \iey{\emph{DRAM banks}}. \iey{DRAM
\emph{cells} in a DRAM bank are laid out in a two-dimensional structure of rows
and columns.} {A} DRAM cell {stores one bit of data} {in the form of} electrical
charge in {a} capacitor, which can be accessed through an access transistor. A
wire called \omcr{2}{\emph{wordline}} drives the gate of all DRAM cells' access
transistors in a DRAM row. A wire called \emph{bitline} connects all DRAM cells
in a DRAM column to a common differential sense amplifier. Therefore, when a
wordline is asserted, each DRAM cell in the DRAM row is connected to its
corresponding sense amplifier. The set of sense amplifiers is called \emph{the
row buffer}, where the data of an activated DRAM row is buffered to serve a
column access \atbcr{5}{operation}.}

\subsection{DRAM Operation}

\copied{The memory controller serves memory access requests by issuing DRAM
commands, e.g., row activation ($ACT$), bank precharge ($PRE$), data read
($RD$), data write ($WR$), and refresh ($REF$) while respecting certain timing
parameters to guarantee correct operation~\cite{jedec2020lpddr5,
jedec2015lpddr4,jedecddr,jedec2020ddr4,jedec2012ddr3,jedecddr5c,jedec2021hbm}.
{The memory controller issues} an $ACT$ command alongside the bank
address and row address corresponding to the memory request's address
{to activate a DRAM row}. During the row activation process, a
DRAM cell loses its charge, and thus, its initial charge needs to be restored
(via a process called \emph{charge restoration}). The latency from the start of
a row activation until the completion of the DRAM cell's charge restoration is
called the \emph{\gls{tras}}. To access another row in an already activated DRAM
bank, the memory controller must issue a $PRE$ command to close the opened row
and prepare the bank for a new activation.}

\copied{A DRAM cell is inherently leaky and thus loses its stored electrical
charge over time. To maintain data integrity, a DRAM cell {is periodically
refreshed} with a {time interval called the \emph{\gls{trefw}}, which is
typically} \SI{64}{\milli\second} (e.g.,~\cite{jedec2012ddr3, jedec2020ddr4,
micron2014ddr4}) or \SI{32}{\milli\second} (e.g.,~\cite{jedec2015lpddr4,
jedecddr5c, jedec2020lpddr5}) at normal operating temperature (i.e., up to
\SI{85}{\celsius}) and half of it for the extended temperature range (i.e.,
above \SI{85}{\celsius} up to \SI{95}{\celsius}).  
To refresh all cells \omcr{2}{in a timely manner}, the memory controller
{periodically} issues a refresh {($REF$)} command with {a time interval called}
the \emph{\gls{trefi}}, {which is typically} \SI{7.8}{\micro\second}
(e.g.,~\cite{jedec2012ddr3, jedec2020ddr4, micron2014ddr4}) or
\SI{3.9}{\micro\second} (e.g.,~\cite{jedec2015lpddr4, jedecddr5c,
jedec2020lpddr5}) at normal operating temperature. When a rank-level refresh
command is issued, the DRAM chip internally refreshes several DRAM rows
\hpcareva{in all banks}, during which the whole rank is busy.}
\hpcalabel{A5}\hpcareva{When a bank-level refresh command is issued,
\hpcareva{the DRAM chip internally refreshes DRAM rows in several banks, during
which these banks are busy.}}

\subsection{DRAM Read Disturbance}

\copied{Read disturbance is the phenomenon \gra{in which} reading data from a memory or
storage device causes physical disturbance (e.g., voltage deviation, electron
injection, electron trapping) on another piece of data that is \emph{not}
accessed but physically located \gra{near} the accessed data. Two prime examples
of read disturbance in modern DRAM chips are RowHammer~\cite{kim2014flipping}
and RowPress~\cite{luo2023rowpress}, where repeatedly accessing (hammering) or
keeping active (pressing) a DRAM row induces bitflips in physically nearby DRAM
rows. In RowHammer and RowPress terminology, the row that is
hammered or pressed is called the \emph{aggressor} row, and the row that
experiences bitflips the \emph{victim} row. For read disturbance bitflips to
occur, 1)~the aggressor row needs to be activated more than a certain threshold
value, \param{which we call the read disturbance threshold (defined
in~\secref{sec:introduction})}\gra{,} and/or
2)~\omcr{2}{\acrfull{taggon}}~\cite{luo2023rowpress} needs to be large
enough~\cite{kim2020revisiting, orosa2021deeper, yaglikci2022understanding,
luo2023rowpress}.}

\copied{To avoid read disturbance bitflips, systems take preventive
actions, e.g., they refresh victim
rows~\refreshBasedRowHammerDefenseCitations{}, selectively throttle accesses to
aggressor rows~\cite{yaglikci2021blockhammer, greenfield2012throttling}, \gf{or}
physically isolate potential aggressor and victim rows~\cite{hassan2019crow,
konoth2018zebram, saileshwar2022randomized, saxena2022aqua, wi2023shadow,
woo2023scalable}. These solutions aim to perform preventive actions before the
cumulative effect of an aggressor row's \emph{activation count} and \emph{on
time} causes read disturbance bitflips.}

\subsection{Motivation}

\begin{figure}[!ht]
    \centering
    \includegraphics[width=.9\linewidth]{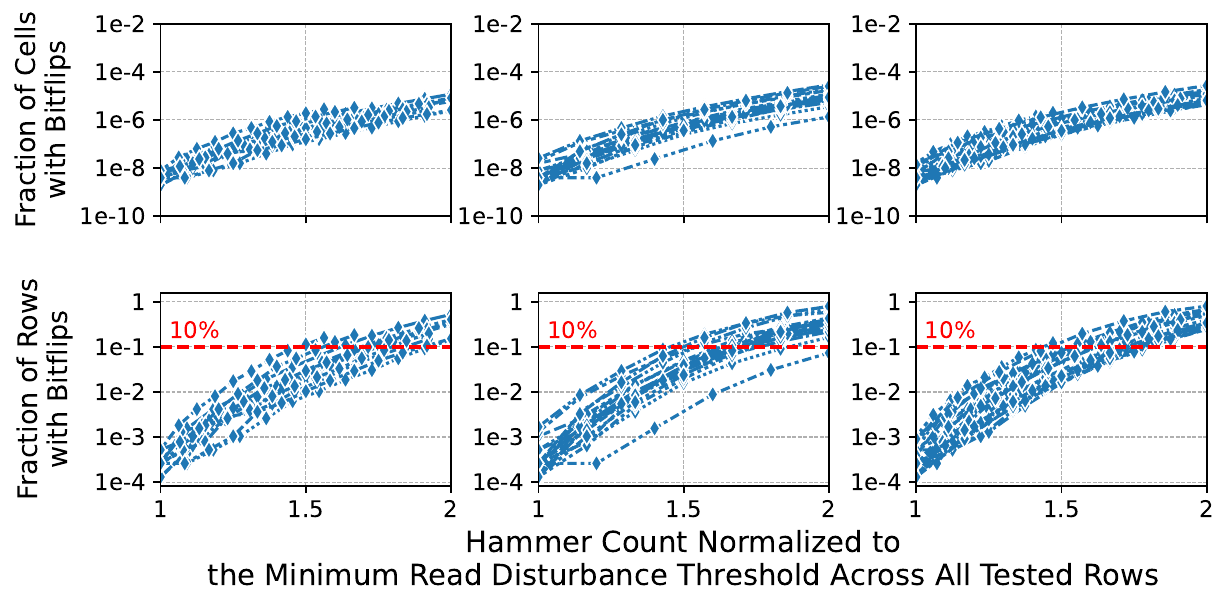}
    \caption{The fraction of tested cells (top) and the fraction of 
    tested rows with bitflips
    (bottom) over varying hammer counts normalized to the minimum read
    disturbance threshold across all tested rows (x-axis) for each tested bank} 
    \label{fig:bitflips_vs_rdt}
\end{figure}

Using the experimental methodology described in detail 
in~\secref{sec:experimental-infra}, we measure the RDT of each tested row
in all tested modules \emph{once}. \figref{fig:bitflips_vs_rdt} shows the
fraction of tested cells with a bitflip (top) and fraction of tested rows with
a bitflip (bottom) across modules from 
each tested major manufacturer (left, middle, right) over hammer count (x-axis).
We display hammer count values normalized to \gls{minrdt}.

Two important observations from this figure motivate our work. First, \emph{only} 
a small fraction of DRAM rows exhibit the smallest RDT values in a chip. 
Thus, \gls{minrdt} likely falls below the read disturbance 
threshold of a large fraction of DRAM rows, even though the
RDT of a row changes over time~\cite{olgun2025variable}. It is important
to understand how reliable a read disturbance mitigation technique
configured after one or few measurements of \gls{minrdt} is. This is because measuring
\gls{minrdt} one or few times is relatively less time-intensive than measuring
the RDT of all DRAM rows hundreds of thousands of times to account for
temporal variation in RDT.
Second, a large fraction (90\%) of rows exhibit larger RDT values 
than 1.5-2.0$\times{}$ \gls{minrdt}. 
This observation is in line with prior 
work~\cite{orosa2021deeper,yaglikci2024spatial}.
If DRAM rows that exhibit smaller RDT than 1.5-2.0$\times{}$ \gls{minrdt}
can be reliably and rapidly identified, spatial variation-aware
read disturbance mitigation techniques (i.e., Svärd~\cite{yaglikci2024spatial}) 
could become widely adopted.
\section{Experimental Infrastructure}
\label{sec:experimental-infra}

We describe our DRAM testing infrastructure and the real DDR4 DRAM
chips tested.

\begin{figure}[ht]
\centering
\includegraphics[width=\linewidth]{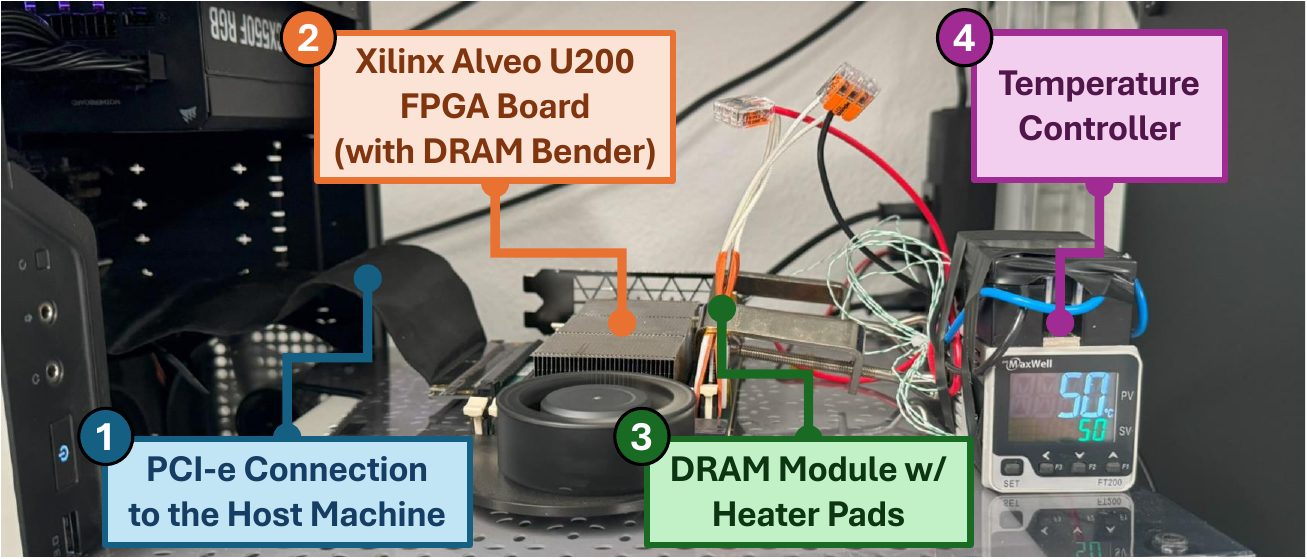}
\caption{Our DRAM Bender~\cite{olgun2023drambender} based experimental setup}
\label{fig:infra}
\end{figure}

\begin{figure}[ht]
\centering
\includegraphics[width=\linewidth]{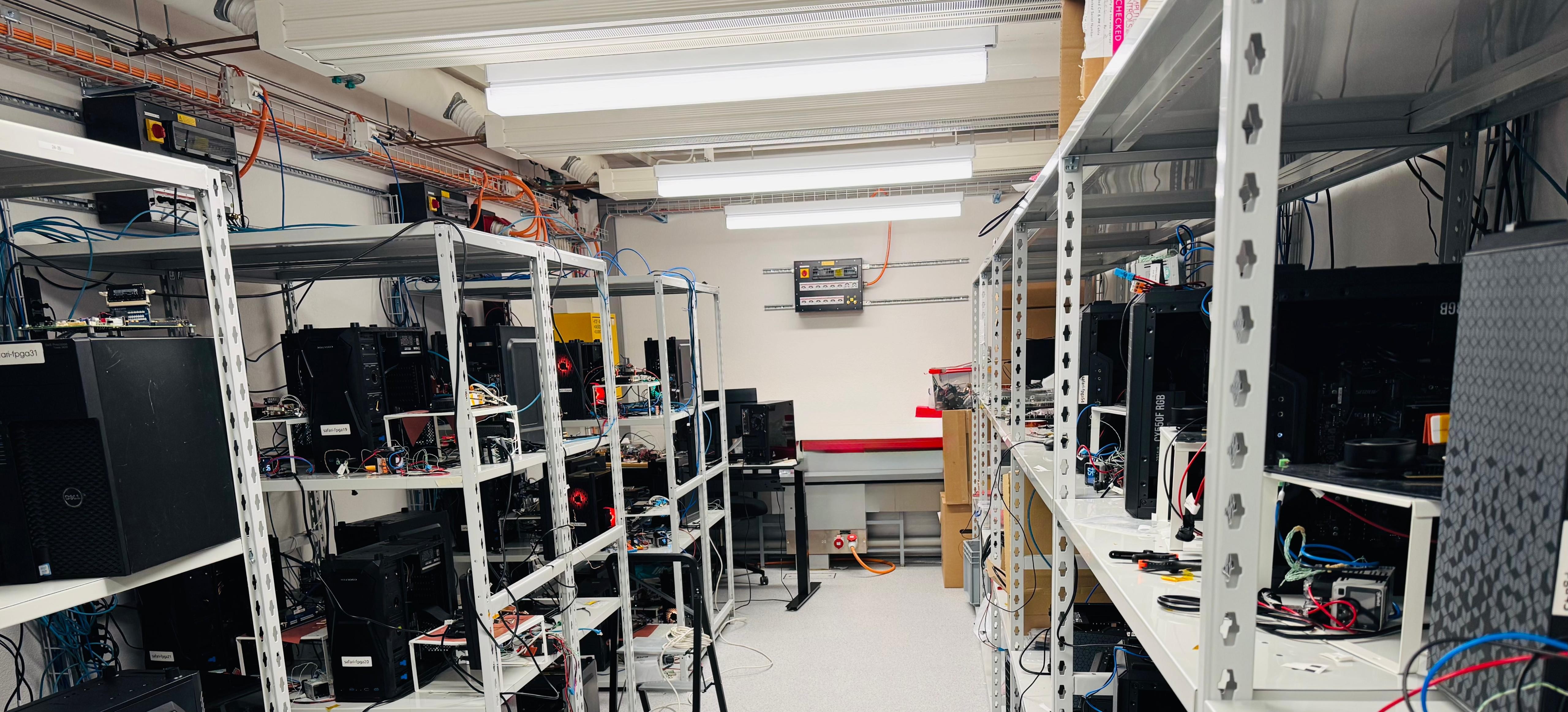}
\caption{Our laboratory for real DRAM chip experiments}
\label{fig:lab}
\end{figure}

\noindent
\textbf{DRAM Testing Setup.} We build our testing setup on DRAM
Bender~\cite{olgun2023drambender,safari-drambender}, an open\gra{-}source
FPGA-based DRAM testing infrastructure \omcr{2}{(which builds on
SoftMC~\cite{hassan2017softmc, softmc-safarigithub})}. \figref{fig:infra} shows one of our experimental DDR4
module testing setups, which consists of
four parts \iey{1})~a computer that executes test programs and collects
experimental results, \iey{2})~an Alveo U200~\cite{alveo-u200} FPGA development
board for DDR4 modules, programmed with DRAM Bender to execute test programs
\agy{and analyze experimental data}, 3)~a thermocouple\agy{-based} temperature
sensor and a pair of heater pads pressed against the DRAM chips that {heat} up
the DRAM chips to a desired temperature, and 4)~a PID temperature controller
(MaxWell FT200~\cite{maxwellFT200}) that keeps the temperature at \atbcr{4}{the}
desired level with a precision of $\pm$\SI{0.5}{\celsius}. \figref{fig:lab} shows our
laboratory comprising many DDR4 testing platforms.

\noindent
\textbf{Tested DRAM Chips.}
Table~\ref{tab:dram_chip_list} shows the {\numDDRchips{} DDR4 DRAM chips (in
\nummodules{} modules) that we test from all three major DRAM manufacturers. {To
study \X{}'s sensitivity to different DRAM technologies, designs, and
manufacturing processes, we test various} DRAM chips with different densities,
die revisions, chip organizations, and DRAM standards.

\begin{table}[!ht]
  \centering
  \footnotesize
  \caption{Tested DDR4 Modules}
  \begin{adjustbox}{max width=\linewidth}
    \begin{tabular}{c|ccccc}
\multicolumn{1}{l|}{}                                                        & \textbf{DDR4}        & \textbf{\# of}       & \textbf{Density}     & \textbf{Chip}        & \textbf{Date}         \\
\multicolumn{1}{l|}{\textbf{DRAM Mfr.}}                                           & \textbf{Module}      & \textbf{Chips}       & \textbf{Die Rev.}    & \textbf{Org.}        & \textbf{ww-yy}        \\ 
\hline\hline
\multirow{4}{*}{\begin{tabular}[c]{@{}c@{}}Mfr. H\\ (SK Hynix)\end{tabular}} & M1                   & 8                    & 16Gb – C             & x8                   & 52-23                 \\
                                                                             & M2                   & 8                    & 16Gb – C             & x8                   & 36-21                 \\
                                                                             & M3, M4, M5           & 8                    & 8Gb – D              & x8                   & 20-24                 \\
                                                                             & M6                   & 8                    & 8Gb – D              & x8                   & 38-19                 \\ 
\hline
\multirow{6}{*}{\begin{tabular}[c]{@{}c@{}}Mfr. M\\ (Micron)\end{tabular}}   & M7                   & 8                    & 16Gb – B             & x8                   & 29-22                 \\
                                                                             & M8, M9               & 4                    & 16Gb – E             & x16                  & 46-20                 \\
                                                                             & M10, M12, M13        & 8                    & 16Gb – F             & x8                   & 12-24                 \\
                                                                             & M11                  & 8                    & 16Gb – F             & x8                   & 37-22                 \\
                                                                             & M14                  & 8                    & 8Gb – R              & x8                   & 10-24                 \\
                                                                             & M15, M16             & 8                    & 8Gb – R              & x8                   & 12-24                 \\ 
\hline
\multirow{8}{*}{\begin{tabular}[c]{@{}c@{}}Mfr. S\\ (Samsung)\end{tabular}}  & M17                  & 8                    & 16Gb – A             & x8                   & 20-23                 \\
                                                                             & M18, M19             & 8                    & 16Gb – A             & x8                   & 02-23                 \\
                                                                             & M20, M21             & 8                    & 16Gb – B             & x8                   & 15-23                 \\
                                                                             & M22                  & 4                    & 4Gb – C              & x16                  & 19-19                 \\
                                                                             & M23                  & 8                    & 8Gb – C              & x8                   & N/A                   \\
                                                                             & M24, M25             & 8                    & 16Gb – C             & x8                   & 08-24                 \\
                                                                             & M26, M27             & 8                    & 8Gb – D              & x8                   & 10-21                 \\
                                                                             & M28                  & 8                    & 16Gb – M             & x8                   & N/A                   \\ 
\hline\hline
\multicolumn{1}{l}{}                                                         & \multicolumn{1}{l}{} & \multicolumn{1}{l}{} & \multicolumn{1}{l}{} & \multicolumn{1}{l}{} & \multicolumn{1}{l}{} 
\end{tabular}
    \end{adjustbox}
    \label{tab:dram_chip_list}
    \vspace{-5mm}
\end{table}

\subsection{Testing Methodology}
\label{subsec:testing_methodology}

\noindent
\textbf{Disabling Sources of Interference.}
We identify \param{three} \omcr{2}{factors} that can interfere with our results:
1)~\agy{data retention failures}~\cite{liu2013experimental, patel2017reaper},
2)~on-die read disturbance defense mechanisms (e.g.,
TRR~\cite{frigo2020trrespass, hassan2021utrr,micron2018ddr4trr}), 
and 3)~\agy{error correction codes
(ECC)}~\cite{jedec2021hbm,patel2020beer,patel2021harp}. We carefully reuse the
state-of-the-art read disturbance characterization methodology used in prior
works to eliminate the interference \omcr{2}{factors}~\cite{kim2020revisiting,
orosa2021deeper, yaglikci2022understanding, hassan2021utrr, luo2023rowpress,
yaglikci2024spatial,olgun2023hbm, olgun2024read}. First, we make sure that our
experiments finish strictly within \agy{a} refresh window, in which the DRAM
manufacturers guarantee \agy{that \emph{no}} retention bitflips
occur~\cite{jedec2020ddr4, jedec2021hbm}.
Second, \agy{we disable} periodic refresh \agy{as doing so} disables all known
on-die read disturbance defense
mechanisms~\cite{orosa2021deeper,yaglikci2022understanding,
kim2020revisiting,hassan2021utrr}. 
Third, we {verify} that the tested DDR4 chips {do
\emph{not} have on-die ECC}~\cite{patel2020beer, patel2021harp}, and {we do
\emph{not} use rank-level ECC in our testing setup.}

\noindent
\copied{\textbf{RowHammer Access Pattern}. We use double-sided RowHammer~\cite{kim2014flipping,kim2020revisiting,orosa2021deeper,
seaborn2015exploiting, luo2023rowpress}, which alternately activates two
aggressor rows \omcr{2}{physically adjacent to} a victim row. We record the
bitflips observed in the row between two aggressor rows.}

\noindent 
\copied{\textbf{Logical-to-Physical Row Mapping}. DRAM manufacturers use mapping
schemes to translate logical (memory-controller-visible) addresses to physical
row addresses~\cite{kim2014flipping, smith1981laser, horiguchi1997redundancy,
keeth2001dram, itoh2013vlsi, liu2013experimental, seshadri2015gather,
khan2016parbor, khan2017detecting, lee2017design, tatar2018defeating,
barenghi2018software, cojocar2020rowhammer,  patel2020beer,
yaglikci2021blockhammer, orosa2021deeper}. To identify aggressor rows that are
physically adjacent to a victim row, we reverse-engineer the row mapping scheme
following the methodology described in prior work~\cite{orosa2021deeper}.} 

\noindent
\textbf{RowHammer Test Parameters}. We perform multiple different tests with
varying test parameters. First, \emph{\omcr{2}{Hammer count}}: We define the
\emph{hammer count} of a double-sided read disturbance access pattern as the
number of activations \emph{each} aggressor row receives. Therefore,
\atbcr{4}{in} a double-sided RowHammer test with a hammer count of 5, we
activate each of the two aggressor rows 5 times, resulting in a total of 10 row
activations. Second, \emph{Aggressor row on time}: The time each aggressor row
stays \omcr{2}{open after} each activation during a RowHammer test. We set
aggressor row on time to \SI{35}{\nano\second} (the minimum aggressor row on
time as defined in the DDR4 interface specification). Third, \emph{Data
pattern}: we use a single data pattern to initialize the victim (V), the
aggressors (V$\pm$1), and the 6 DRAM rows surrounding each aggressor row
(V$\pm$[2:8]). We initialize the victim row with \emph{all-0s} aggressor rows
with \emph{all-1s}, and the surrounding rows with \emph{all-0s}.~\iey{We use this
data pattern as several real DRAM characterization works~\cite{luo2023rowpress,
luo2024experimental,luo2025revisiting} demonstrate that the dominant bitflip
direction for RowHammer is 0 to 1.} Fourth, \emph{Temperature}: We use a
temperature controller setup for all DDR4 modules and set the target temperature
to \SI{50}{\celsius} \hpcarevcommon{unless explicitly mentioned}.}

\noindent
\textbf{Read Disturbance Threshold.} 
We quantify \agy{the read disturbance vulnerability of a DRAM row using the}
\emph{\omcr{2}{read disturbance threshold (RDT)}} \omcr{2}{metric, i.e.,}
the \emph{hammer count needed to induce the first read disturbance bitflip} in
the victim row.
\section{Major Results}
\label{sec:foundational_results}

We study the variation in read disturbance threshold (RDT) across DRAM rows and
repeated measurements. We measure RDT in all tested rows (\numtestedrows{} such
rows \hpcalabel{C2}\hpcarevc{in total across all tested modules}) in DDR4 DRAM chips 1,000
times. 
\outline{Motivate this study that we are doing to see how min RDT changes.}
\outline{Any key result to summarize here?}

\noindent
\textbf{DRAM Testing Algorithm.} 
\hpcalabel{C2}\hpcarevc{We test a fraction
($1/16$) of all DRAM rows in each tested DRAM bank. 
We randomly
determine the addresses of the tested DRAM rows \emph{once} and we keep testing these
rows across all experiments in our work.} 
First,
we identify \gls{minrdt} by hammering
every tested victim row varying number of times using the minimum \gls{taggon}
(e.g., \SI{35}{\nano\second}). To quickly identify the \gls{minrdt}, we sweep
the hammer count from 1,000 to 25,000 in increments of 1,000.
The first step yields the $RDT_{min}$ value as the
minimum read disturbance threshold across all tested rows. Second,
we repeatedly measure the RDT of all tested rows. We
do so by testing the DRAM row for read disturbance failures using hammer counts
ranging from $RDT_{min}/2$ to $RDT_{min}*2$ with increments of
$RDT_{min}/30$.\footnote{We {empirically} determine the range and the
granularity of the tested hammer count values such that our experiments take
reasonable time and cover a wide range of RDTs that the DRAM row may exhibit.}
This experiment yields a series of 1,000  
RDT measurements for each tested row.

\begin{figure}[!ht]
    \centering
    \includegraphics[width=.95\linewidth]{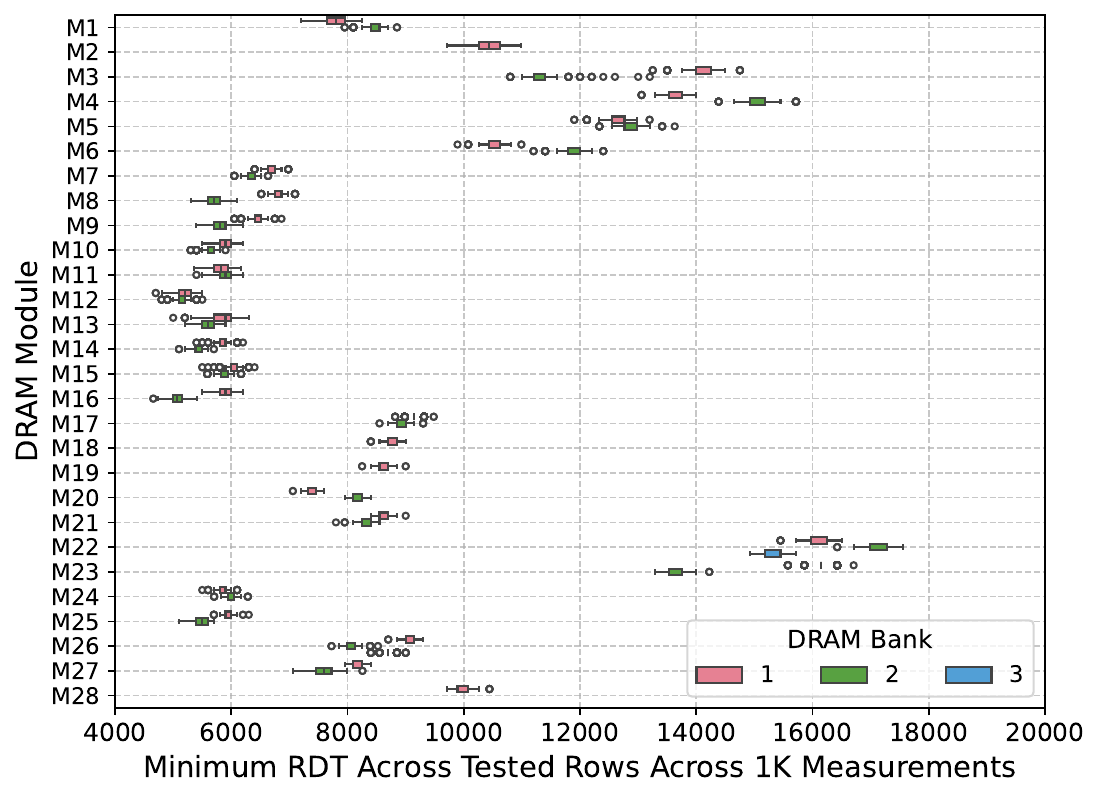}
    \caption{\gls{minrdt} across all measurements 
    for each tested DDR4 module}
    \label{fig:minimum_rdt_across_rows_across_measurements}
\end{figure}

\figref{fig:minimum_rdt_across_rows_across_measurements} shows the \gls{minrdt} distribution observed in each of the 1,000 measurements for every
tested DRAM module in a box-and-whiskers plot.
Each box shows the
distribution of values in one bank. \hpcalabel{A5}\hpcareva{Some distributions have no boxes
drawn (e.g., M23 bank 1). For such distributions, the majority of RDT measurements
yield the same value as each other.}

We make \param{two} major observations from
\figref{fig:minimum_rdt_across_rows_across_measurements}. First, \gls{minrdt} in
a bank varies across repeated measurements (over time). For example, in bank 1
of M1 (the box at the top of the figure) the \gls{minrdt} is \param{7200} in the
\param{100}th measurement and \param{8250} in the \param{182}nd measurement. Second,
\gls{minrdt} in a DRAM bank varies across repeated measurements in all tested
banks of DDR4 modules, that is, no tested DRAM bank exhibits the
same \gls{minrdt} across 1,000 measurements.

\take{\gls{minrdt} varies with repeated measurements (over time) in all tested
DRAM banks.}

\begin{figure}[!ht]
    \centering
    \includegraphics[width=.85\linewidth]{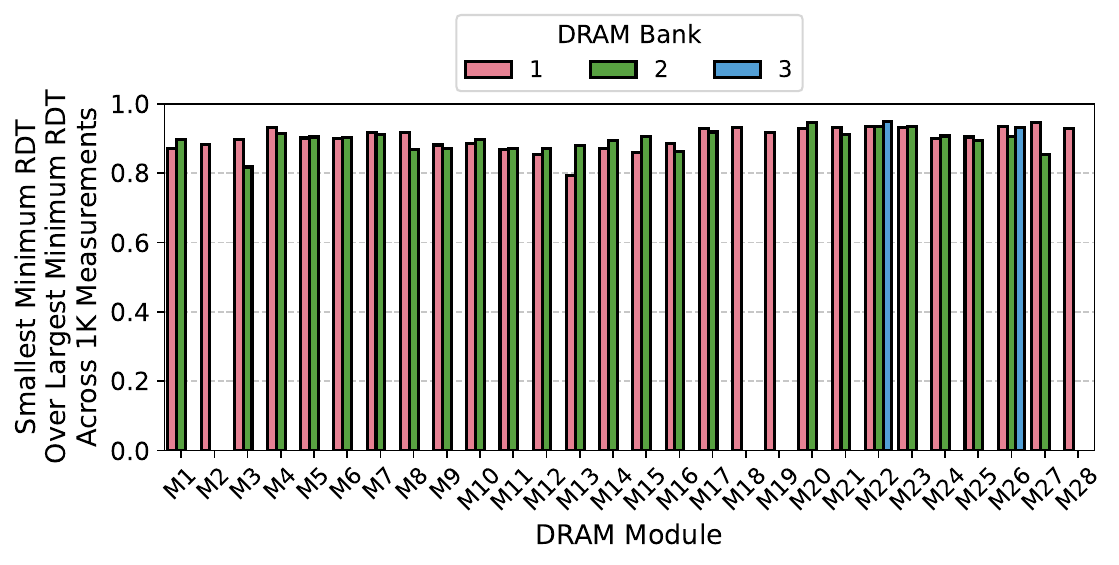}
    \caption{The smallest \gls{minrdt} across measurements normalized to the
    largest across measurements (y-axis) for each tested DDR4 module
    (x-axis)}
    \label{fig:max_normalized_rdt}
\end{figure}

\figref{fig:max_normalized_rdt} quantifies the degree of the variation in
\gls{minrdt} across measurements. Each bar shows the minimum \gls{minrdt}
divided by the maximum \gls{minrdt} values across all measurements for a tested
DRAM bank.

From \figref{fig:max_normalized_rdt}, we observe that \gls{minrdt} varies
significantly over time. The minimum \gls{minrdt} across measurements is
\param{9.9\%}, \param{5.1}\%, and \param{21.0}\% smaller than the maximum
\gls{minrdt} across measurements on average, at least, and at most across all
tested DRAM banks, respectively. This indicates that a \emph{guardbanding-based
solution} for read disturbance threshold identification (e.g., measuring
\gls{minrdt} once and reducing it by a factor to hopefully cover all possible
read disturbance bitflips across time) requires at least a \param{21.0}\%
guardband.\hpcalabel{A5}\footnote{\hpcareva{While we have extensively characterized the read
disturbance threshold of 317985 rows using 1000 test iterations, we \emph{cannot}
guarantee that 21.0\% guardband would prove robust for all DRAM chips that are
in use today. Therefore, a robust guardbanding solution for RDT 
identification likely requires a guardband larger than 21.0\%.}}

\begin{figure}[!ht]
    \centering
    \includegraphics[width=.85\linewidth]{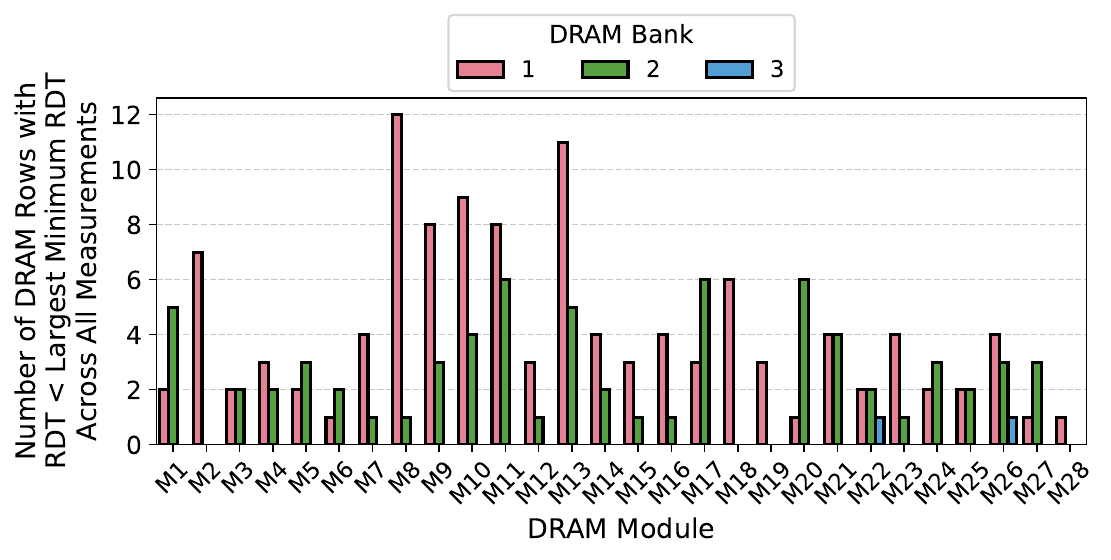}
    \caption{The number of DRAM rows that exhibit read disturbance bitflips when
    the largest minimum read disturbance threshold across all measurements is
    used as the hammer count}
    \label{fig:unique_row_counts}
\end{figure}

\figref{fig:unique_row_counts} shows the number of unique DRAM row addresses
that exhibit at least one read disturbance bitflip across 1,000 measurements,
when we use a hammer count $\leq$ maximum~\gls{minrdt}\ieycomment{max(rdtmin)} across 1,000
measurements.

We make two important observations from \figref{fig:unique_row_counts}. First,
multiple different DRAM rows may exhibit read disturbance bitflips at hammer
counts below the largest\ieycomment{highest or maximum} \gls{minrdt} across 1,000 measurements. In M8, as many as 12 DRAM rows exhibit smaller read disturbance
thresholds than the largest \gls{minrdt}. These DRAM rows are difficult to
identify using a single RDT measurement. Second, across 1,000 measurements,
several tested DRAM banks have \emph{only} one DRAM row that yields \gls{minrdt}
(e.g., \param{bank 1 in M28}). Such a row is relatively easy to identify using a
single RDT measurement.

\take{\gls{minrdt} changes significantly over time. Many DRAM rows can exhibit
read disturbance bitflips at RDT $\leq$ the largest \gls{minrdt} across 1,000
measurements.}\ieycomment{you already have the first sentence in takeaway 1}

\begin{figure}[!ht]
    \centering
    \includegraphics[width=\linewidth]{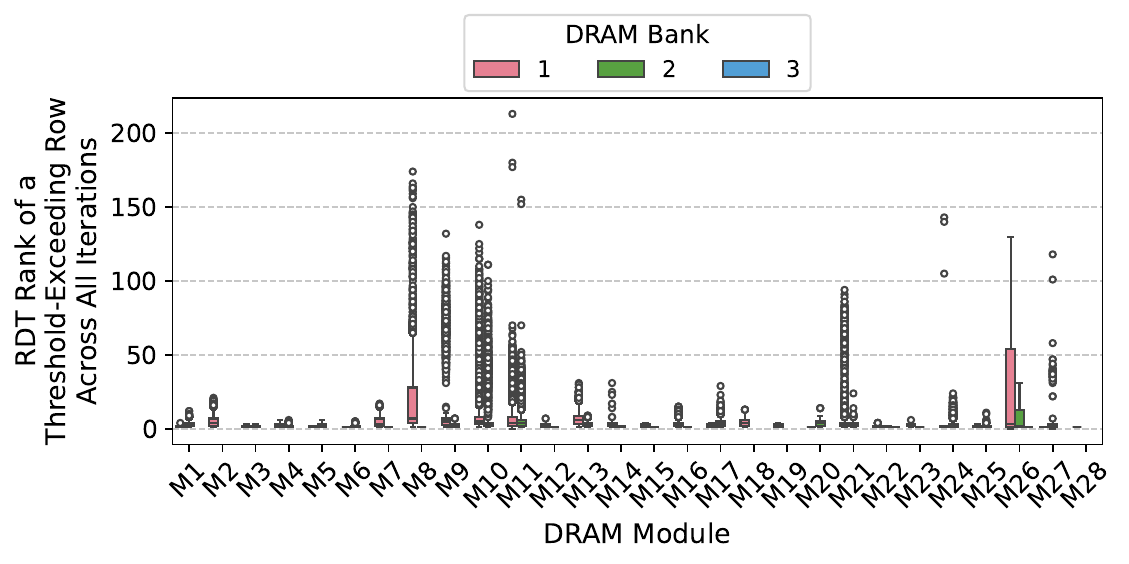}
    \caption{The RDT rank of a DRAM row that exhibits a read disturbance bitflip
    when the largest minimum read disturbance threshold is used as the hammer
    count. A higher value indicates that a DRAM row can have both a relatively
    high and a very low RDT across measurements.}
    \label{fig:row_placement_across_iterations}
\end{figure}

While there are \emph{only} at most tens of DRAM rows whose RDTs fall below the
largest \gls{minrdt} across 1,000 measurements in a bank, these DRAM rows are
\emph{not} necessarily the rows with the smallest RDTs in every measurement. We
sort the tested DRAM rows in increasing RDT order for each measurement.
\figref{fig:row_placement_across_iterations} shows the distribution of the rank
(in that sorted order) of each DRAM row whose RDT falls below the largest
\gls{minrdt} across 1,000 measurements. 

From \figref{fig:row_placement_across_iterations}, we observe that a DRAM row
can be among the rows with the smallest few RDT values in one measurement, and
can also have a relatively high RDT value in another measurement. For example, a
row in \param{bank 1 in M11} has the smallest RDT value across all rows for
\param{498} measurements and has the 213th smallest RDT value for one
measurement. This indicates that identifying all potential DRAM rows that can
fall below a certain read disturbance threshold value (e.g., with the goal of
disabling such rows to prevent all read disturbance bitflips), even for very low
read disturbance values (e.g., \gls{minrdt}), is challenging.

\take{The DRAM row with the \gls{minrdt} in one measurement can have a very
large RDT value (e.g., the 213th smallest RDT across all tested rows in a bank)
in another measurement.}\ieycomment{large RDT distribution? large and value sounds weird} 

According to our empirical read disturbance bitflip dataset for hammer count
$\leq$ maximum \gls{minrdt} across all measurements (\param{187} such rows),
applying a \param{21.0\%} safety margin to the \gls{minrdt} obtained by
measuring the read disturbance threshold of all DRAM rows \emph{only once}
prevents all observed bitflips. However, making a general conclusion about the
bitflip prevention effectiveness of using a safety margin for the read
disturbance threshold is challenging and requires more data and large-scale
experimental studies~\cite{olgun2025variable}. Thus, a practical and secure
approach to mitigating read disturbance bitflips likely requires some error
tolerance (e.g., error-correcting codes) because 1)~testing for read disturbance
bitflips is time- and energy-intensive 
and
2)~one read disturbance threshold measurement is unlikely to yield the minimum
\gls{minrdt} across time.

\section{\X{}: Towards Practical and Reliable Read Disturbance Testing}
\label{sec:probabilities}
Securely preventing read disturbance bitflips in a system requires carefully
configuring the read disturbance threshold of the employed read disturbance
mitigation technique: if the threshold is set too low, the mitigation technique
induces unnecessary and even prohibitive system performance and energy
overheads~\cite{olgun2024abacus,canpolat2024understanding,
yaglikci2021blockhammer,kim2020revisiting,qureshi2022hydra,
canpolat2025chronus,yaglikci2024spatial}\nbcomment{comet?}, whereas if the threshold is set higher than the minimum
\gls{minrdt} across time, some read disturbance bitflips become unavoidable.
Moreover, it is difficult to definitively identify an \gls{minrdt} that falls
below all RDT values that any DRAM row in a DRAM chip might exhibit.

Because there is no absolutely reliable approach to identifying the minimum
\gls{minrdt} in a DRAM chip across time, understanding the reliability of a read
disturbance threshold identification methodology that uses \emph{only} one RDT
measurement for each DRAM row is a potential approach to developing a robust RDT
identification methodology. To that end, using our empirical read disturbance
bitflip dataset, we analyze the read disturbance bitflip probability for a
worst-case scenario whereby we pick the \emph{largest \gls{minrdt}} across 1,000
measurements \hpcareve{across all tested rows} and repeatedly hammer all tested
victim DRAM rows \gls{minrdt} times. 

\begin{figure}[!ht]
    \centering
    \includegraphics[width=.85\linewidth]{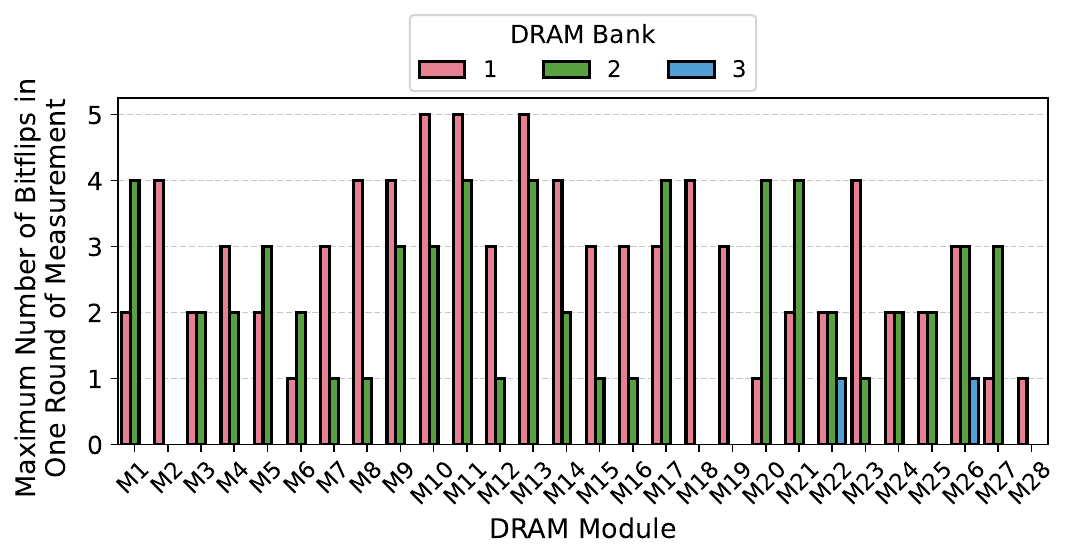}
    \caption{The maximum number of read disturbance bitflips across 1,000 
    test iterations. In each test iteration, all tested DRAM rows are 
    hammered with hammer count = the largest \gls{minrdt} across 1,000
    measurements.}
    \label{fig:bitflips_per_iteration}
\end{figure}

\figref{fig:bitflips_per_iteration} shows the maximum number of bitflips we
uncover across 1,000 test iterations (in each iteration, we hammer each tested
row for \emph{largest \gls{minrdt}} \hpcalabel{E4}\hpcareve{(across all rows and
across 1000 measurements)} times). \nbcomment{this sort of conflicts with the
last sentence of the previous paragraph}

We observe that if the \gls{minrdt} value identified through one RDT measurement
is the largest across 1,000 measurements (and the read disturbance techniques
are configured using this value),\nbcomment{this seems redundant and confusing with the previous previous paragraph.} a DRAM module can exhibit up to (as few as)
\param{5} (\param{1}) bitflips when all tested\ieycomment{why parantheses in tested? didnt get it} DRAM rows are each hammered
with a hammer count that equals the identified RDT value.

\take{Measuring \gls{minrdt} once and using it as the read disturbance threshold
can lead to multiple read disturbance bitflips.}

Without error tolerance, these bitflips \emph{cannot} be prevented solely by a
read disturbance mitigation technique. We make the key observation that while
there can be as many as \param{5} bitflips (when the one-time measurement of
\gls{minrdt} is used to configure the mitigation technique), these bitflips are
distributed to a select few rows in each tested module
\hpcalabel{A4}(\hpcareva{i.e., the bitflip distribution encompasses various
rows, the bitflips are not confined to a single row, see}
\figref{fig:unique_row_counts}). Thus, with some form of error tolerance, these
bitflips could be mitigated. 

\noindent
\textbf{Modeling Bitflip Probability.} To quantify the effectiveness of
combining error tolerance with read disturbance mitigation configured using a
one-time \gls{minrdt} measurement in preventing read disturbance bitflips, we
develop an empirically-driven model for read disturbance 
bitflips.\hpcalabel{B1}\footnote{\hpcarevb{We 
do not know enough about the underlying causes of the
temporal variation in read disturbance threshold and we believe that it should
be studied independently of our work. Therefore, we experimentally study hundreds of 
DRAM chips to develop and evaluate our fast and 
reliable RDT identification methodology. We hope and expect that future
device-level studies (inspired by this work) will develop a better understanding
of the inner workings of the temporal variation in read disturbance threshold.
}} The model has two important
components: 1)~possible bitflip locations ($L$), 2)~number of bitflips ($N$). We
conservatively assume that the possible bitflip locations grow at a constant
rate ($\Delta{}L$)\hpcalabel{A1}\footnote{\hpcareva{We consider the linear
growth of $L$ to be conservative because the rate of new
(previously-not-observed) bitflip locations reduce with test iterations (across
1,000 test iterations), similar to what is shown in
Figure~\ref{fig:spatial_new_unique_bitflips}.}} \hpcarevc{depicted
in}~\figref{fig:unique_row_counts} and the number of bitflips is the maximum
empirically encountered in a test iteration as depicted
by~\figref{fig:bitflips_per_iteration}. 

\hpcalabel{C3}At every 1,000 epochs, $L$ grows by $\Delta{}L$ uniform randomly
distributed bitflip locations across the simulated DRAM
chip.
\hpcarevc{That is, $L$ is the set of unique locations (bit addresses) that could
possibly harbor bitflips and $\Delta{}L$ is the number of new unique (uniform
randomly distributed) locations that are added to $L$ every 1,000 epochs.} At
every epoch (corresponding to the time it takes to hammer all tested DRAM rows),
we evaluate the model by uniform randomly selecting $N$ locations from $L$.
\hpcarevc{In other words, $N$ is the number of bitflips that are encountered by
the model in every epoch and these bitflips can appear \emph{only} in 
possible bitflip locations stored in $L$.}
\hpcarevc{We explain how we determine $\Delta{}L$ and N for a tested DRAM bank
over an example DRAM module M8 (bank 1). $\Delta{}L$ for M8 is 12
from~\figref{fig:unique_row_counts} and $N$ for M8 is 5
from~\figref{fig:bitflips_per_iteration}.}

\begin{figure}[!ht]
    \centering
    \includegraphics[width=.9\linewidth]{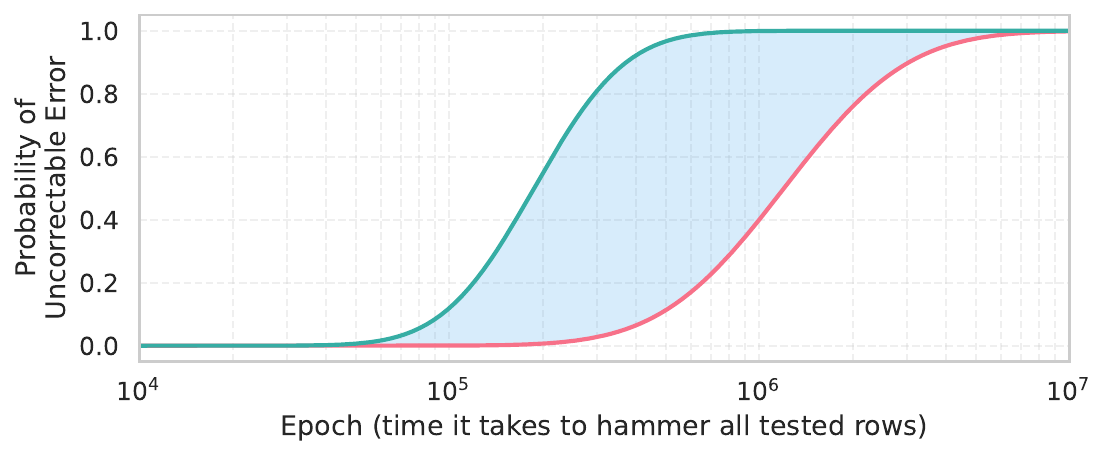}
    \caption{Probability of uncorrectable error over epochs based on the
    simulated read disturbance bitflip model. All simulated modules yield
    curves that fall between the pink and green curves in the shaded area.}
    \vspace{-2mm}
    \label{fig:failure_probability}
\end{figure}

Using this model, we evaluate the probability of uncorrectable error as two or
more bitflips manifesting in a 136-bit error-correcting code
(ECC)~\cite{dell1997white, gong2018memory, hamming1950error, kang2014co,
meza2015revisiting, micron2017whitepaper, nair2016xed, patel2019understanding,
biancasigmetrics09, dram-field-analysis3} codeword possibly implemented in DDR5
chips today~\cite{mineshphd, alam2022comet, kim2023unity} for area efficiency
and low cost overhead. We run 10K Monte Carlo simulations and evaluate the
uncorrectable error probability for varying epochs.
\figref{fig:failure_probability} shows the simulated probability of
uncorrectable error (y-axis) over epochs (x-axis).\footnote{We assume that the
read disturbance bitflips induced in an epoch are corrected at the end of 1000
epochs. This can be done with relatively infrequent ECC scrubbing operations.
E.g., double-sided hammering a victim row 10K times takes
\SI{900}{\micro\second} and doing that in a bank of 256K rows 1K times takes
approximately 65 hours. The overheads of scrubbing a DRAM bank every 65 hours is
negligible and can be largely alleviated with in-DRAM ECC scrubbing techniques
(e.g.,~\cite{hassan2024self,jedecddr5c}). We leave uncorrectable error
probability analysis with varying ECC scrubbing intervals for future work.} The
green and pink curves each represent the worst- and best-case uncorrectable
error probabilities derived from read disturbance bitflip models based on
individual modules.

We observe that the probability of uncorrectable errors largely vary between the
simulations (based on empirical data observed from the tested modules). We
compute the mean time to an uncorrectable error (MTTUE) in terms of number of
epochs based on the two curves as \param{216248}\ieycomment{is this hour? what is the metric here} (green curve) and
\param{1560087} (pink curve).\footnote{We exclude two modules that yield a zero
probability of uncorrectable error from this analysis. Since $N$ for these two
modules are 1, our model does \emph{not} predict any uncorrectable errors for
these two modules.} These MTTUE values are relatively low. For example, given
there are 256K rows in a bank and the hammer count = 10K, hammering all rows in
a bank takes approximately 0.07 hours, yielding an MTTUE of \emph{only}
\param{1.42e+04} and \param{1.02e+05} hours for the green and pink curves,
respectively. 
\outline{Add one sentence comparing this to empirical retention error MTTFs}
We attribute the
relatively low read disturbance MTTUE (i.e., relatively high uncorrectable error
rate) to 1)~the conservative assumptions we made throughout the analysis steps
and 2)~the modeled workload access pattern that repeatedly hammers all DRAM rows
in the bank. E.g., a workload access pattern that does \emph{not} always hammer
all DRAM rows in a bank could lead to a significantly higher MTTUE. We leave the
design and evaluation of less conservative read disturbance bitflip models
(building on \X{}) for future work.

\take{Measuring \gls{minrdt} once and using it as the read disturbance threshold
for all DRAM rows lead to a relatively high uncorrectable error
probability.}

\begin{figure}[!ht]
    \centering
    \includegraphics[width=.9\linewidth]{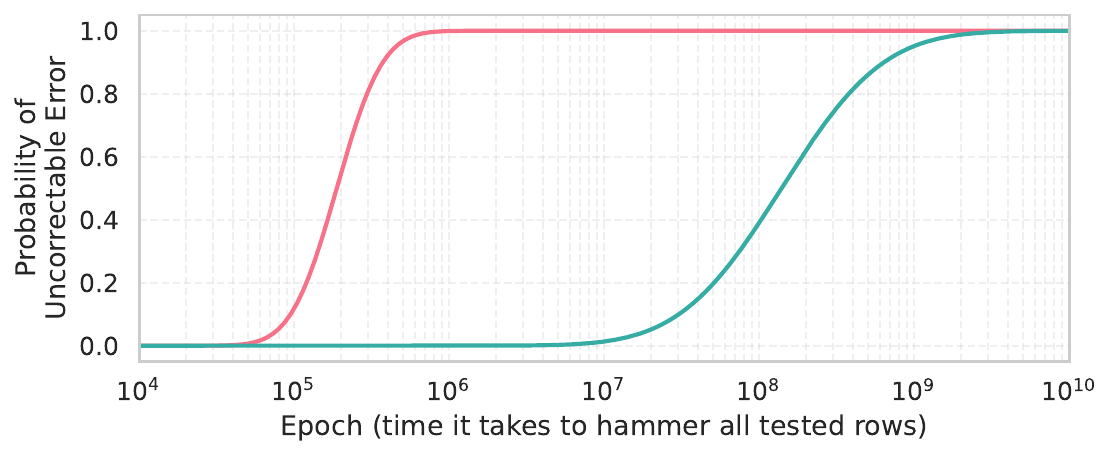}
    \caption{\iey{Probability of uncorrectable error over epochs for when a cell
    that exhibits a read disturbance bitflip is detected and removed from the
    set of possible bitflip locations (the green curve on the right). The pink
    curve depicts the probability of error for the worst-case module
    in~\figref{fig:failure_probability}.}}
    \label{fig:failure_probability_new}
\end{figure}

To curb the uncorrectable error probability, a system designer could decide to
apply a very large safety margin (e.g., 50\%) to the configured read disturbance
threshold of a DRAM row after it is found to exhibit bitflips. Doing so would
keep the set of possible bitflip locations ($L$) as small as possible.
\figref{fig:failure_probability_new} (green curve on the right) shows the
probability of uncorrectable error over epochs in case the possibility of read
disturbance bitflips in a DRAM row is eliminated after a bitflip in the row is
detected. We plot the probability for the module that yields the worst-case
uncorrectable error probability in~\figref{fig:failure_probability}. The pink
curve depicts the probability of uncorrectable error
shown in~\figref{fig:failure_probability}.

We observe that the probability of uncorrectable error over time reduces
significantly. For example, for the modeled worst-case DRAM module, 
mean time between uncorrectable errors improves from \param{1.42e+04} hours to
\param{1.84e+07} hours. 

\take{The read-disturbance-induced uncorrectable error probability can be
significantly reduced by eliminating the possibility of a bitflip manifesting in
a DRAM row (e.g., by applying a large safety margin to the RDT) after the row
exhibits the first read disturbance bitflip.}

\subsection{\hpcarevcommon{\X{}'s Sensitivity to Temperature}}

\hpcalabel{Common Concern \#2}\hpcarevcommon{Variable read disturbance is
affected by temperature~\cite{olgun2025variable}. Therefore, we investigate
\X{}'s sensitivity to temperature.}

\begin{figure}[!ht]
    \centering
    \includegraphics[width=\linewidth]{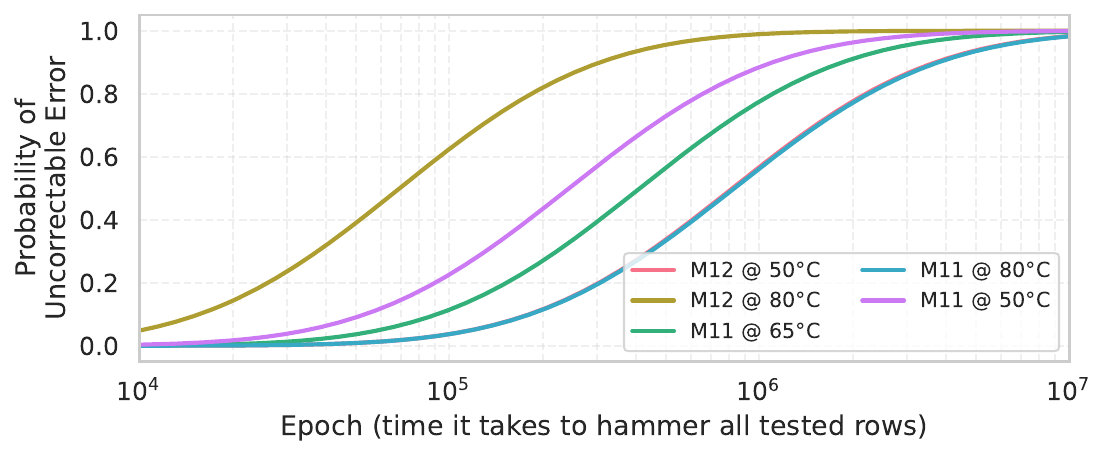}
    \caption{\hpcarevcommon{Probability of uncorrectable error over epochs for
    three tested temperature levels}}
    \label{fig:temperature_error_probability}
\end{figure}

\hpcarevcommon{We observe that the key parameters of our error probability model
can significantly change with temperature. 
\figref{fig:temperature_error_probability} shows the uncorrectable
error probability (in case the possibility of a read disturbance bitflip in a
DRAM row is eliminated after a bitflip in the row is detected) we compute using
our error model for all tested temperature and DRAM module combinations.}

\hpcarevcommon{We observe that the error probability significantly varies with
temperature. For example, for M12 @ \SI{80}{\celsius}, the MTTUE is 8.90e+03 and
for M12 @ \SI{50}{\celsius}, the MTTUE is 1.07e+05.\footnote{\hpcarevcommon{We
do not plot M12 @ \SI{65}{\celsius}
in~\figref{fig:temperature_error_probability}. Empirical results for M12 @
\SI{65}{\celsius} yield a $\Delta{}L$ and $N$ of 1 and 1, respectively.
Therefore, our model predicts that read disturbance bitflips will be correctable
in this module at this temperature according to our empirical results.}}} 

\subsection{\hpcareve{\X{}'s Sensitivity to Aggressor Row On Time}}

\begin{figure}[!ht]
    \centering
    \includegraphics[width=\linewidth]{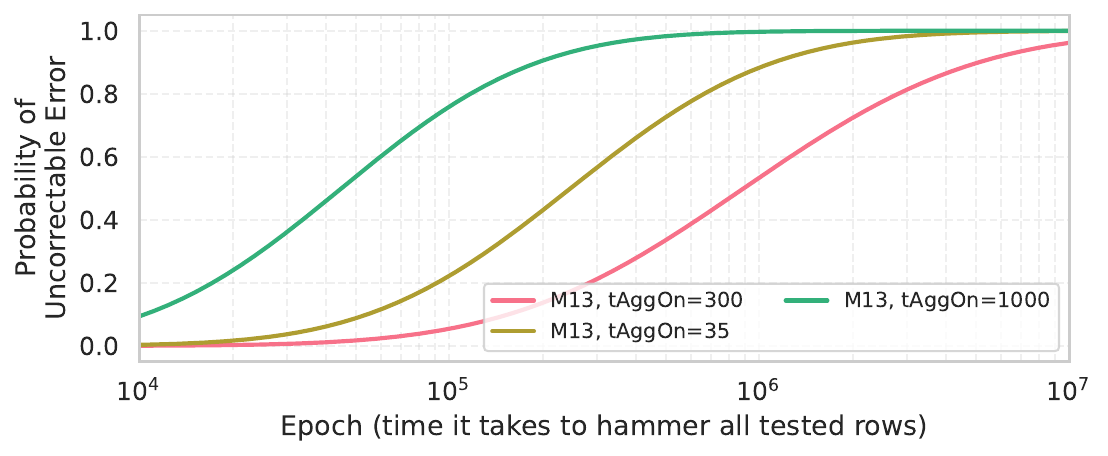}
    \caption{\hpcareve{Probability of uncorrectable error over epochs for
    three tested tAggOn values}}
    \label{fig:taggon_error_probability}
\end{figure}

\hpcalabel{E3}
\hpcareve{We investigate \X{}'s sensitivity to aggressor row on time
(RowPress~\cite{luo2023rowpress}).} 
We observe that \X{}'s key parameters significantly change with tAggOn.
\hpcareve{\figref{fig:taggon_error_probability} shows the error probability for
the three tested tAggOn values. We make the key observation that the MTTUE
ranges from 5.62e+03 (tAggOn=1000ns) to 1.48e+05 (tAggOn=300ns) for a tested
DRAM module (M13).}

\take{\hpcarevcommon{\X{} can be used to quantitatively understand the
sensitivity of a read disturbance mitigation mechanism's reliability to
temperature and aggressor row on time.}}

\section{Spatial Variation Aware Read Disturbance Threshold Characterization}

Applying a large safety margin to ``weak'' DRAM rows that exhibit bitflips (when
hammer count $\leq$ maximum \gls{minrdt} across 1,000 measurements is used to
configure a read disturbance mitigation technique) can greatly reduce the
uncorrectable error probability and thereby improve system robustness against
read disturbance bitflips. However, doing so requires a flexible substrate for
read disturbance mitigation techniques that enables run-time configurable read
disturbance thresholds at DRAM row granularity. Designing that substrate and
integrating it into a DRAM-based computing system is out of the scope of this
paper, but such a substrate would likely be built on the ideas of Spatial
Variation-Aware Read Disturbance Defences (Svärd)~\cite{yaglikci2024spatial}
because~\cite{yaglikci2024spatial} enables configuring the read disturbance of
each DRAM row individually and exploiting the RDT variation across DRAM rows in
a chip for improving read disturbance mitigation performance
(\figref{fig:bitflips_vs_rdt} depicts that variation across rows in the DRAM
banks we test).

We investigate the distribution of the RDT values across tested DRAM rows in a
bank and identify that only a small fraction of DRAM rows (approximately 10\%)
exhibit RDT values below 1.5$\times{}$-2.0$\times{}$ \gls{minrdt}. That is, for
approximately 90\% of DRAM rows, a read disturbance threshold value sufficiently
larger than \gls{minrdt} could be used to reduce the overheads of read
disturbance mitigations~\cite{yaglikci2024spatial}. This observation is in line
with prior work~\cite{orosa2021deeper,yaglikci2024spatial}.\atbcomment{Is it?}
However, the RDT of a DRAM row varies over time~\cite{olgun2025variable} and
determining the read disturbance profile of a DRAM bank using a single
measurement could incorrectly be incorrect. 

We experimentally study the quantity and the physical locations of the read
disturbance bitflips that manifest when a single RDT measurement is used for the
large majority (approximately 90\%) of the DRAM rows. We evaluate the
uncorrectable error probability of a spatial and temporal variation aware read disturbance mitigation mechanism.

\begin{figure}[!ht]
    \centering
    \includegraphics[width=\linewidth]{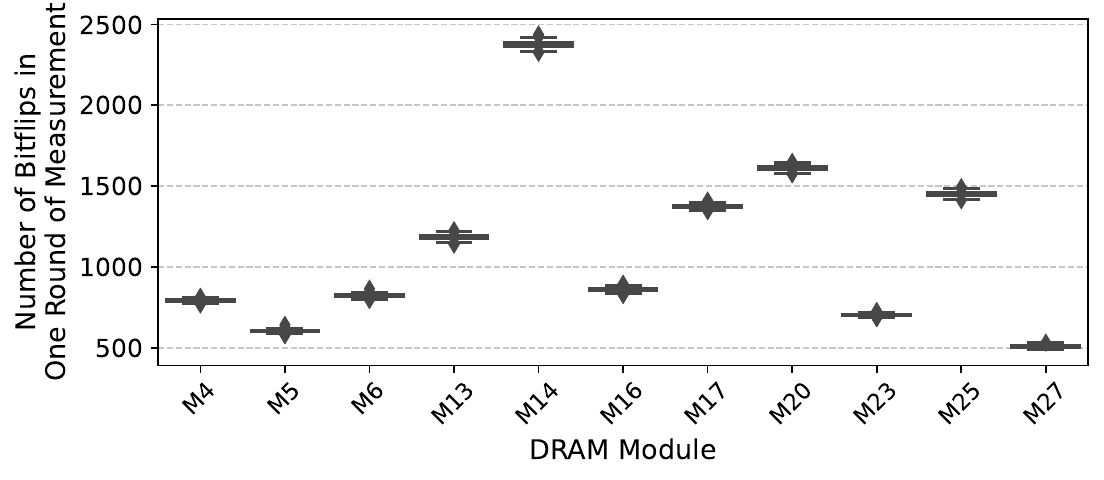}
    \caption{The number of bitflips in each round of measurement across 1K
     repetitions (y-axis) for tested DRAM modules (x-axis).}
    \vspace{-2mm}
    \label{fig:spatial_bitflips_per_iteration}
\end{figure}

\noindent
\textbf{Experimental Methodology.}
From \figref{fig:bitflips_vs_rdt}, we identify an RDT value, $RDT_{90\%}$, that
is smaller than the RDT of 90\% of tested rows for each module (the intersection
of the dashed red line and each plotted curve over the x-axis). We hammer all
tested DRAM rows using hammer count = $RDT_{90\%}$ and collect bitflip
locations. We repeat this procedure 1K times.
\figref{fig:spatial_bitflips_per_iteration} depicts the number of bitflips
across 1K repetitions.

\atbcomment{regenerate all figures with correct module labels}

We make two observations from
\hpcalabel{A5}\figref{fig:spatial_bitflips_per_iteration}. First, there are many
bitflips across all tested rows when tested DRAM rows are repeatedly hammered
$RDT_{90\%}$ times. For example, there are as many as \param{2437} bitflips
across \param{3854} tested DRAM rows in module \param{M14}. Second, the
variation in the number of bitflips across test repetitions is small for all
tested DRAM modules. We attribute this variation to variable read
disturbance~\cite{olgun2025variable}.

\take{Hammer count = $RDT_{90\%}$ yields many read disturbance bitflips.}

\begin{figure}[!ht]
    \centering
    \includegraphics[width=\linewidth]{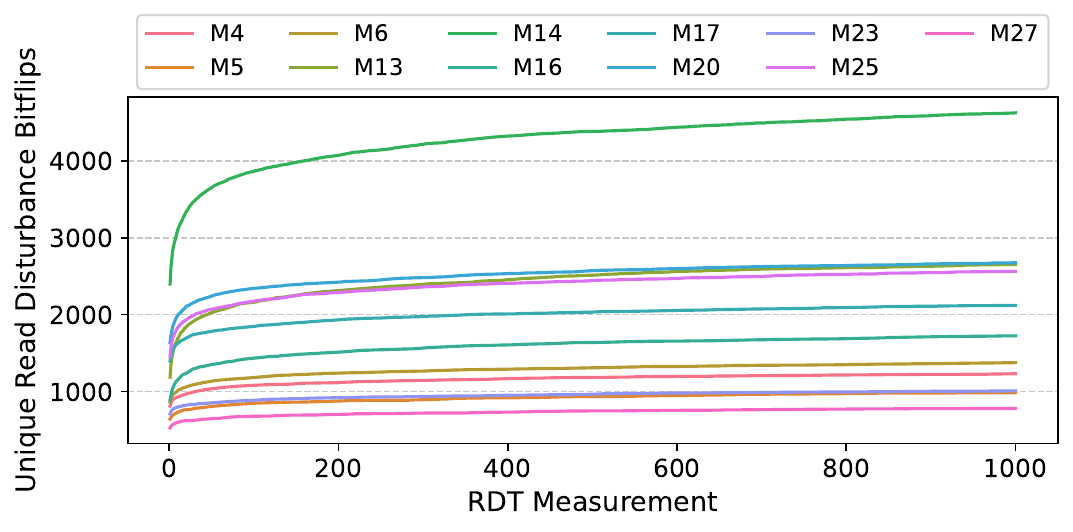}
    \caption{\iey{Unique read disturbance bitflips across repeated RDT
    measurements (x-axis) for various DRAM modules.}}
    \label{fig:spatial_cumulative_unique_bitflips}
\end{figure}

To show how quickly such bitflips appear across test repetitions, we plot the
number of unique bitflip locations discovered in the DRAM chip over test
repetitions in \figref{fig:spatial_cumulative_unique_bitflips}.

We make a major observation from
\figref{fig:spatial_cumulative_unique_bitflips}. The number of unique bitflips
significantly increase with more test repetitions. A considerable fraction of
bitflips \emph{cannot} be identified with a single RDT measurement alone. For
example, for M14, \param{48.1\%} of bitflip locations appear after the first RDT
measurement. We observe similar trends across all tested modules.

\begin{figure}[!ht]
    \centering
    \includegraphics[width=\linewidth]{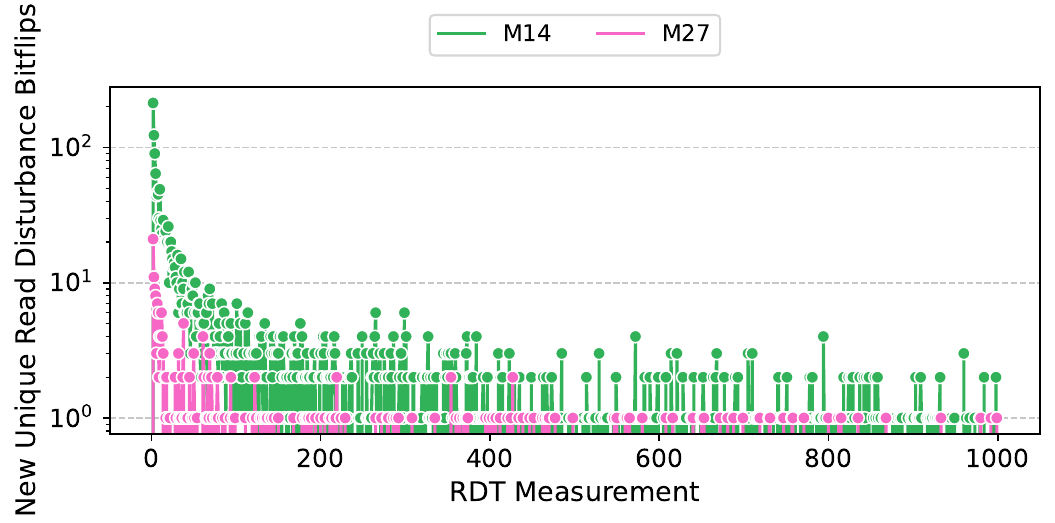}
    \caption{\iey{Newly discovered unique read disturbance bitflips across
    repeated RDT measurements (x-axis) for two representative DRAM modules.}}
    \vspace{-2mm}
    \label{fig:spatial_new_unique_bitflips}
\end{figure}

\figref{fig:spatial_new_unique_bitflips} shows the number of new read disturbance
bitflip locations discovered with each test repetition.

We observe that the number of new read disturbance bitflip locations steadies
with repeated measurements. For example, towards the end of 1K test repetitions,
only one or two new bitflip locations are discovered for the two example
modules.

\take{Repeated RDT measurements are needed to uncover a large fraction of all
observed read disturbance bitflip locations. The rate at which new locations are
discovered tends to slow down and approach 1 as test repetitions approach 1K.}

We posit from our analysis that a spatial variation-aware approach to read
disturbance mitigation (e.g., Svärd~\cite{yaglikci2024spatial}) requires
multiple rounds of read disturbance testing to uncover the large majority of
bitflips that appear due to temporal variation in the RDT
of a DRAM row. If not uncovered, these bitflips could easily overwhelm the
widely-used error tolerance techniques' (e.g., single-error correcting
double-error detecting codes~\cite{kim2015bamboo} or Chipkill
ECC~\cite{locklear2000chipkill,dell1997white,jian2013adaptive}) error correction
capabilities and
violate memory isolation, degrade application quality, or deny
service\exploitingRowHammerAllCitations{}.\ieycomment{check line 103-104 in tex.}

Using the read disturbance bitflip modeling methodology in
\secref{sec:foundational_results}, we evaluate the uncorrectable error
probability for a Svärd-like approach. We assume that the system designer
determines $RDT_{90\%}$ and performs tens of test repetitions using hammer
count = $RDT_{90\%}$ to uncover a large fraction of the VRD-induced
bitflips.\hpcalabel{A5}\footnote{Performing so many read disturbance test
repetitions would take a very long time~\cite{olgun2025variable}. If test
repetitions are performed all at once, the test time could be prohibitive.
However, test repetitions could be distributed across time using online error
profiling techniques (e.g.,~\cite{patel2017reaper}\hpcareva{)}.} We compute the
probability of uncorrectable error for a range of $N$ = $\Delta{}L$ that overlap
our empirical observations.

\begin{figure}[!ht]
    \centering
    \includegraphics[width=\linewidth]{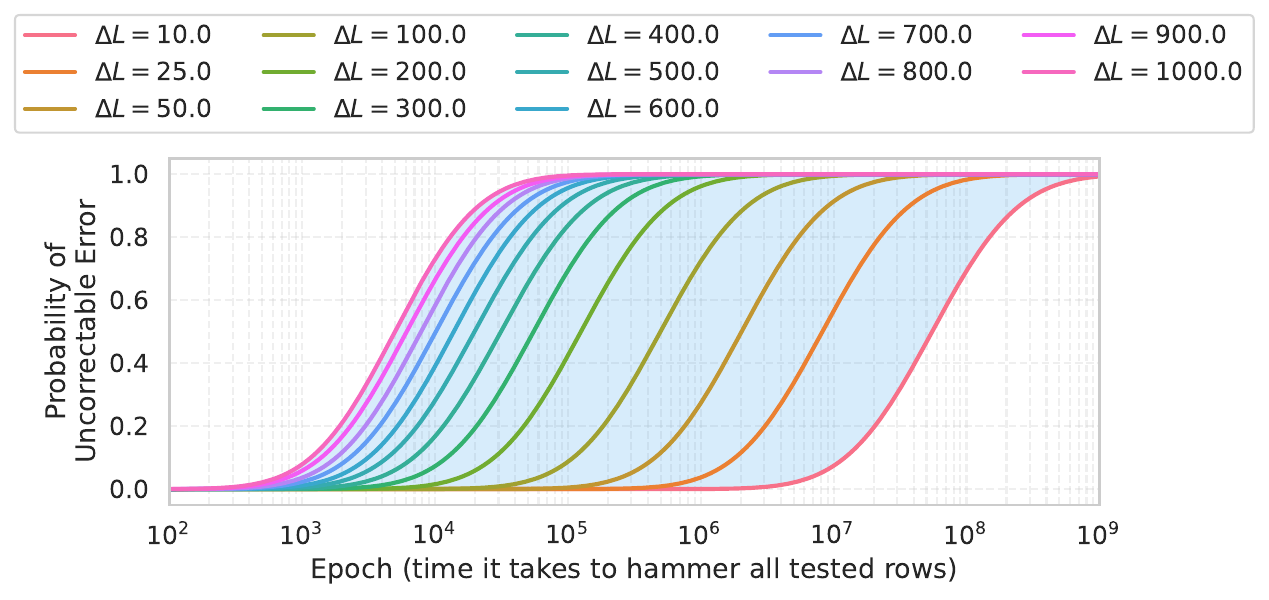}
    \caption{\iey{Probability of uncorrectable errors over epochs for Svärd 
    for varying $\Delta{}L$ values}}
    \label{fig:spatial_failure_probability}
\end{figure}

We make two major observations. First, a high $\Delta{}L$ of 1000 (every epoch,
a new bitflip location is discovered) the uncorrectable error probability is
significantly high for even small number of epochs. This probability yields a
Mean Time to Uncorrectable Error (MTTUE) of \emph{only} \param{6.23e+02} hours.
Second, smaller $\Delta{}L$ values (e.g., 10, on average one new bitflip
location is discovered every 100 epochs) yield smaller uncorrectable error
probability and higher MTTUE. For example, $\Delta{}L$ = 10 yields an MTTUE of
\param{7.25e+06} hours.

\take{Empirically observed bitflip distributions yield relatively high
uncorrectable error probability. An uncorrectable error can manifest as
frequently as every \param{6.23e+02} hours. To reduce error probability,
repeatedly testing for read disturbance failures is important.}

\noindent
\textbf{Effect of ECC Scrubbing Frequency.}
The uncorrectable error probability can be reduced with more frequent ECC
scrubbing operations. The key idea is to discover read disturbance bitflips
before multiple such bitflips accumulate in one ECC codeword. At a relatively
high ECC scrubbing frequency (e.g., approximately 3.9 minutes for a bank of 256K
rows, when a hammer count of 10K is used to hammer each DRAM row once), a
scrubbing operation can complete in one epoch. Such high-frequency scrubbing
operations could discover all empirically observed bitflips for a subset of the
evaluated modules whose number of new read disturbance bitflip locations
(see~\figref{fig:spatial_new_unique_bitflips}) converges to \emph{only} one in
an epoch. We leave detailed exploration of the error probability -- ECC
scrubbing frequency tradeoff for future work.

\noindent
\textbf{A Spatial and Temporal Variation-Aware Read Disturbance Mitigation
Technique.} We observe that the uncorrectable error probability for a 
Svärd-like approach to read disturbance mitigation is similar to 
the error probabilities for a \emph{one size fits all} 
approach to read disturbance mitigation (see~\secref{sec:probabilities}). 
Building on our empirical results and analyses, we propose a
methodology for configuring the read disturbance threshold of Svärd. This way,
we exploit the spatial variation in read disturbance threshold across DRAM rows
to improve system performance and energy efficiency. We also quantify the
uncorrectable error probability (as presented earlier in this section) given
temporal variation in read disturbance. Thereby, we provide quantitative
insights into system robustness against read disturbance bitflips for Svärd.

We evaluate a version of Svärd that enables dynamically configuring a DRAM row's
RDT at runtime. We use two distinct RDT values:
1)~\gls{minrdt} with a safety margin of 21\% subtracted from it (such a value
would mitigate all experimentally observed bitflips in this study) and
2)~$RDT_{90\%}$ that is used as the RDT of a majority of DRAM rows to perform
fewer read disturbance mitigative actions (e.g., refreshing the victim rows of
an aggressor row) and improve system performance. Following initial read
disturbance threshold profiling using $RDT_{90\%}$, we expect 20\% of DRAM rows
to fall below the $RDT_{90\%}$ and we configure our Svärd model to reflect
the empirical RDT distribution across rows.

\noindent
\textbf{\hpcareva{Evaluation Methodology.}}\hpcalabel{A2} We use \param{57}
single-core workloads from SPEC CPU2006~\cite{spec2006}, SPEC
CPU2017~\cite{spec2017}, TPC~\cite{tpcweb}, MediaBench~\cite{fritts2009media},
and YCSB~\cite{ycsb} to construct \param{60} four-core workload mixes.
\hpcareva{Based on the row buffer misses-per-kilo-instruction (RBMPKI)
{metric}~\cite{yaglikci2021blockhammer}, we group single-core workloads into
three categories: 1) low memory-intensity (L), RBMPKI $\in [0,2)$, 2) medium
memory-intensity (M), RBMPKI $\in [2,10)$, 3) high memory-intensity (H), RBMPKI
$\in [10+)$, by analyzing {each workload's} SimPoint~\cite{calder2005simpoint}
traces (200M instructions). We construct four-core workload mixes with
increasing memory intensities (from LLLL to HHHH) by randomly selecting four
single-core workloads from each group.} We construct 10
workloads each for each of the following RBMPKI mixes: LLLL, LLHH, MMLL, MMMM,
HHMM, HHHH. Chronus (in-DRAM)~\cite{canpolat2025chronus} has storage overhead
and tracks the activation count of an aggressor row using hardware counters and
preventively refreshes \omcr{2}{the victim rows} before the
activation count reaches the \omcr{3}{configured} read disturbance threshold.
PARA (memory-controller-based)~\cite{kim2014flipping} does \emph{not} have
storage overhead and determines the target row of a DRAM activate command as an
aggressor row \atbcr{2}{based on a probability} \omcr{3}{that is determined
based on the configured RDT} and preventively refreshes \atbcr{2}{the aggressor
row's} neighbors.
 
\begin{figure}[!ht]
    \centering
    \includegraphics[width=\linewidth]{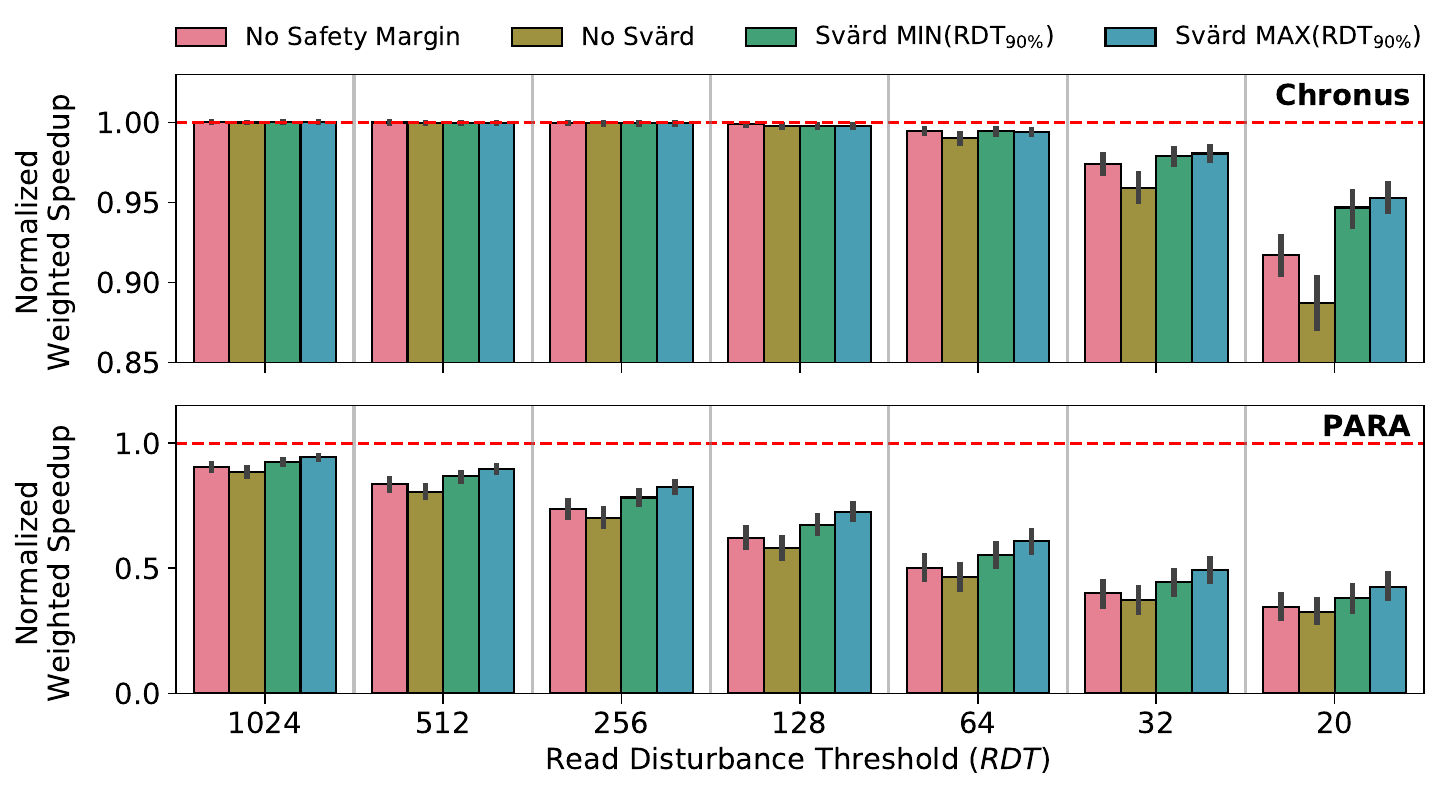}
    \caption{System performance (y-axis, higher is better) for Chronus (PRAC) and PARA
    when spatial variation-aware read disturbance thresholds are used}
    \label{fig:spatial_performance_improvement}
\end{figure}

\copied{\figref{fig:spatial_performance_improvement} shows system performance
\omcr{2}{with} \nb{Chronus (top) and PARA (bottom) alone} and combined with
Svärd, normalized to the baseline system that does \emph{not} implement read
disturbance mitigation using \param{60} four-core  workload
mixes.} \nb{We evaluate Chronus
and PARA without Svärd in two configurations: 1) "No Safety Margin" shows the
mitigation technique configured with \emph{no} safety margin (i.e., VRD agnostic
and potentially unsafe), and 2) "No Svärd" shows the mitigation technique
configured with the safety margin (i.e., with 79\% of $RDT_{min}$). We annotate
each configuration with Svärd in the form of Svärd-[Selected $RDT_{90\%}$],
where the selected $RDT_{90\%}$ is the highest and lowest $RDT_{90\%}$ values
observed across all tested modules.} \nb{The x-axis shows \omcr{2}{seven}
\atbcr{3}{different} worst-case read disturbance threshold values ($RDT$) from
1024 down to 20 to evaluate system performance for modern and
future DRAM chips as technology node scaling over generations exacerbates read disturbance vulnerability.}

\iey{We make two key observations. First,}
\nb{configuring Svärd's read disturbance threshold dynamically with our
empirical results and analyses improves system performance for all tested read
disturbance thresholds compared to \textit{No Safety Margin} and \textit{No
Svärd} configurations. For example, at $RDT$=32, Svärd with the highest
$RDT_{90\%}$ (with the lowest $RDT_{90\%}$) improves the average system
performance with PARA by \param{32\%} (\param{18\%}) compared to \textit{No
Svärd} and by \param{22\%} (\param{10\%}) compared to \emph{No Safety Margin}.
Performance benefits are relatively smaller for Chronus at most tested $RDT$
values compared to PARA. This is because Chronus' performance overhead at these
$RDT$ values is already very low \hpcalabel{A3}\hpcareva{(i.e., even without Svärd).
Therefore, a spatial-variation aware Chronus provides substantial performance
benefits over Chronus only at very low read disturbance thresholds (e.g.,
$\le{}$ 64).} At \textit{RDT}=20, Svärd with the highest $RDT_{90\%}$ (with the
lowest $RDT_{90\%}$) improves system performance by \param{8\%} (\param{6\%})
compared to \emph{No Svärd}, and by \param{4\%} (\param{3\%}) compared to
\emph{No Safety Margin}.}

\nb{Second, only employing a safety margin by configuring the mitigation
techniques with 79\% of $RDT_{min}$ results in performance overheads compared to
the potentially unsafe \emph{No Safety Margin} configuration for all tested
$RDT$ values. For example, at $RDT$=32, employing the safety margin reduces the
system performance with PARA by \param{8\%} compared to \emph{No Safety Margin}.
The performance overheads are relatively smaller for Chronus at most tested
$RDT$ values. At $RDT$=20, employing the safety margin reduces the system
performance with Chronus by \param{3\%} compared to \emph{No Safety Margin}.}

\take{A spatial and temporal-variation aware read disturbance mitigation
significantly improves system performance compared to a \emph{one size fits all}
approach to read disturbance mitigation.}

\ieycomment{An alternative: Instead of applying a fixed
safety margin to rdtmin uniformly across all rows, making the read disturbance
mechanism both spatial and temporal-aware provides both better security and
significantly higher system performance.}

\section{Related Work}
\copied{To our knowledge, this is the first work to} \iey{1) develop a rapid
read disturbance threshold (RDT) testing methodology and 2) comprehensively
examine the RDT measurement time -- read disturbance bitflip probability
tradeoff}.
\iey{In this section, we discuss other relevant prior work.}

\noindent
\textbf{Experimental Read Disturbance Characterization.} 
\copied{Prior works extensively characterize the RowHammer and RowPress
vulnerabilities in real DRAM chips\readDisturbanceCharacterizationCitations{}.
These works 
demonstrate (using real DDR3, DDR4, LPDDR4, and HBM2 DRAM chips) how a DRAM
chip's read disturbance vulnerability varies with 1)~DRAM refresh
rate~\cite{hassan2021utrr,frigo2020trrespass,kim2014flipping}, 2)~the physical
distance between aggressor and victim
rows~\cite{kim2014flipping,kim2020revisiting,lang2023blaster}, 3)~DRAM
generation and technology
node~\cite{orosa2021deeper,kim2014flipping,kim2020revisiting,hassan2021utrr},
4)~temperature~\cite{orosa2021deeper,park2016experiments}, 5)~the time the
aggressor row stays
active~\cite{orosa2021deeper,park2016experiments,olgun2023hbm,olgun2024read,luo2023rowpress,nam2024dramscope,nam2023xray,
luo2025revisiting}, ~6)~physical location of the victim DRAM
cell~\cite{orosa2021deeper,olgun2023hbm,olgun2024read,yaglikci2024spatial},
7)~wordline voltage~\cite{yaglikci2022understanding}, 8)~supply
voltage~\cite{he2023whistleblower}}, \iey{9) reduced DRAM timing
parameters~\cite{tugrul2025understanding, yuksel2025pudhammer}, and 10)
time~\cite{olgun2025variable}.} \iey{\emph{No} prior read disturbance
characterization study provides a testing methodology that can rapidly determine
a reliable RDT value nor evaluate the performance of such a testing
methodology.}

\noindent
\textbf{System-Level RowHammer Tests.}
\copied{Several
works~\cite{farmani2021rhat,cojocar2020rowhammer,zhang2021bitmine,memtest86}
develop tools or RowHammer tests that aim to identify read disturbance bitflips
in DRAM chips in a computing system.} \iey{\X{} could be integrated into such
tools/tests to determine the reliable RDT rapidly.}

\noindent
\textbf{Retention Failure Profiling.}
\copied{Prior works~\cite{liu2013experimental,qureshi2015avatar,patel2017reaper,
khan2014efficacy,khan2016parbor} advocate and propose methods to efficiently
profile DRAM retention failures that are subject to the variable retention time
(VRT) phenomenon. Our work is inspired by the methods proposed in these works
(especially~\cite{qureshi2015avatar}).
}
\section{Conclusion}
We present \X{}, the first reliable and rapid 
read disturbance testing methodology to discover the read disturbance threshold (RDT)
of DRAM chips. \X{} is built on the insights of a rigorous experimental characterization study comprising 
\numDDRchips{} DDR4 chips.
Using \X{}, we evaluate two approaches to read disturbance testing. 
We show that 1)~relying on \emph{a single} RDT measurement 
and a single error-correcting code is insufficient to prevent 
uncorrectable errors with a high probability and 2)~a spatial- and 
temporal-variation aware read disturbance mitigation mechanism could 
improve system performance while remaining as robust 
against read disturbance bitflips
as a conservative approach
to read disturbance mitigation that uses one RDT value for all DRAM rows.
Our results indicate that
a combination of a variety of error tolerance techniques, such as 
memory scrubbing, error-correcting codes, online error profiling is needed 
to achieve sufficient system robustness against read disturbance bitflips.
We hope that \X{} paves the way for more quantitative 
reasoning about the performance, cost, and reliability 
tradeoffs for RDT testing and read disturbance mitigation techniques.

\section*{Acknowledgments}
We thank the anonymous reviewers of {HPCA 2026 and ISCA 2026} for feedback. 
\omcr{2}{We thank the} SAFARI Research Group members for
{constructive} feedback and the stimulating intellectual {environment.} We
acknowledge the generous gift funding provided by our industrial partners
({especially} Google, Huawei, Intel, Microsoft), which has been instrumental in
enabling the research we have been conducting on read disturbance in DRAM {in
particular and memory systems in
general~\cite{mutlu2023retrospectiveflippingbitsmemory, 
mutlu2013memory, mutlu2025memory, mutlu2024memory, mutlu2023fundamentally, mutlu2025modern}.} This work was in part
supported by the Google Security and Privacy Research Award and the Microsoft
Swiss Joint Research Center.


\bibliographystyle{unsrt}
\bibliography{refs}
\balance
\end{document}